\documentclass[sigconf,nonacm]{acmart}

\usepackage[utf8]{inputenc}
\usepackage{tabularx}
\usepackage{geometry}
\usepackage{enumitem} 
\usepackage{makecell} 

\usepackage{multirow} 
\usepackage{array} 
\usepackage{enumitem} 
\usepackage{float}
\usepackage{caption}
\PassOptionsToPackage{square,numbers}{natbib}

\renewcommand{\arraystretch}{1.4}

\renewcommand\footnotetextcopyrightpermission[1]{}
\begin{document}

\title{Secure User-friendly Blockchain Modular Wallet Design\\Using Android \& OP-TEE}

\author{Seongjin Kim}
\authornote{All authors contributed equally to this research.}
\email{mm0ck3r@korea.ac.kr}
\affiliation{%
  \institution{Korea University}
  \city{Seoul}
  \country{South Korea}
}

\author{Sanguk Yun}
\email{dnwjddld94@korea.ac.kr}
\authornotemark[1]
\affiliation{%
  \institution{Korea University}
  \city{Seoul}
  \country{South Korea}
}

\author{Jungho Jang}
\email{jj030116@korea.ac.kr}
\authornotemark[1]
\affiliation{%
  \institution{Korea University}
  \city{Seoul}
  \country{South Korea}
}


\begin{abstract}
\hspace*{1em}Emerging crypto economies still hemorrhage digital assets because legacy wallets leak private keys at almost every layer of the software stack, from user-space libraries to kernel memory dumps. This paper solves that twin crisis of security and interoperability by re-imagining key management as a platform-level service anchored in ARM TrustZone through OP-TEE. Our architecture fractures the traditional monolithic Trusted Application into per-chain modules housed in a multi-tenant TA store, finally breaking OP-TEE’s single-binary ceiling. A cryptographically sealed firmware-over-the-air pipeline welds each TA set to an Android system image, enabling hot-swap updates while Verified Boot enforces rollback protection. Every package carries a chained signature developer first, registry second so even a compromised supply chain cannot smuggle malicious code past the Secure World’s RSA-PSS gatekeeper. Inside the TEE, strict inter-TA isolation, cache partitioning, and GP-compliant crypto APIs ensure secrets never bleed across trust boundaries or timing domains. The Rich Execution Environment can interact only via hardware-mediated Secure Monitor Calls, collapsing the surface exposed to malware in Android space. End-users enjoy a single polished interface yet can install or retire Bitcoin, Ethereum, Solana, or tomorrow’s chain with one tap, shrinking both storage footprint and audit scope. For auditors, the composition model slashes duplicated verification effort by quarantining blockchain logic inside narrowly scoped modules that share formally specified interfaces. Our threat analysis spans six adversary layers and shows how the design neutralizes REE malware sniffing, OTA injection, and cross-module side channels without exotic hardware . A reference implementation on AOSP exports a Wallet Manager HAL, custom SELinux domains, and a CI/CD pipeline that vet community modules before release. The result is not merely another hardware wallet but a programmable substrate that can evolve at the velocity of the blockchain ecosystem. By welding radical extensibility to hardware-anchored assurance, the platform closes the security-usability gap that has long stymied mass-market self-custody. We posit that modular TEEs are the missing OS primitive for Web3, much as virtual memory unlocked multi-tasking in classical computing. Together, these contributions sketch a blueprint for multi-chain asset management that is auditable, resilient, and poised for global deployment.
\end{abstract}


\maketitle

\section{Introduction}
\subsection{Exchange Hacking Incidents}

\subsubsection{2025 Bybit Hack}

Most recently, in February 2025, the Bybit exchange suffered what is now recorded as the largest single cryptocurrency theft in history, with approximately USD 1.4 billion stolen.\cite{intro_1}Based in Dubai, the global exchange Bybit lost 400,000 ETH within a matter of minutes. Investigations revealed that hackers had skillfully exploited the leakage of private keys from the exchange’s hot wallet.\cite{intro_1} The attackers circumvented security verification mechanisms by inserting fraudulent smart contracts disguised as legitimate wallet addresses into the multi-signature wallet transfer process, thereby diverting the funds to external wallets. The CEO of Bybit promptly acknowledged the breach and announced a bounty program to recover the stolen assets. Following international collaborative investigations, on February 26, the FBI officially identified a North Korean hacking group as responsible for the attack.\cite{intro_1} This incident was particularly alarming because it demonstrated that even wallets perceived to be cold wallets could possess vulnerabilities.\cite{intro_6} In 2025 alone, more than USD 2 billion worth of cryptocurrencies were stolen in various hacking incidents, triggering heightened concern among global regulatory bodies and cybersecurity experts.\cite{intro_2} The Bybit hack severely undermined trust in exchange security, prompted urgent security reviews across numerous platforms, and emphasized the need for more proactive and robust security measures throughout the blockchain industry.
\begin{table*}[ht]
\centering
\begin{tabularx}{\textwidth}{|c|c|X|c|X|X|}
\hline
\textbf{Exchange} & \textbf{Year} & \textbf{Attack Method} & \textbf{Amount Stolen} & \textbf{Estimated USD Loss (at the time)} \\
\hline
Bybit & 2025 & Private key leak from hot wallet, forged smart contract & 400,000 ETH & approx. 1.4 billion USD \\
\hline
DMM Bitcoin & 2024 & Wallet system breach (suspected Lazarus Group) & 4,502.9 BTC & approx. 308 million USD \\
\hline
FTX & 2022 & Insider access or abuse of internal privileges & approx. 477 million USD in crypto & approx. 477 million USD \\
\hline
Coincheck & 2018 & Phishing + malware, hot wallet theft & Not disclosed & approx. 534 million USD \\
\hline
Mt. Gox & 2014 & System vulnerabilities and lack of version control & Not disclosed & approx. 460 million USD \\
\hline
\end{tabularx}
\caption{Top 5 Cryptocurrency Exchange Hacking Incidents (as of 2025)}
\end{table*}
\subsubsection{2024 DMM Bitcoin Hack}

In May 2024, Japan’s mid-sized cryptocurrency exchange DMM Bitcoin suffered a major security breach resulting in the theft of 4,502.9 BTC (approximately USD 308 million).\cite{intro_1} The exchange, operated by the IT conglomerate DMM Group since 2018, had been known for its relatively conservative operations. However, hackers managed to penetrate the wallet system and siphon off large amounts of Bitcoin. Investigations pointed strongly toward the North Korean Lazarus Group as the perpetrators, with stolen funds being laundered through corporate accounts in third-party countries.\cite{intro_5} In response, DMM Bitcoin secured emergency funding of approximately USD 320 million to stabilize operations and replenish customer holdings.\cite{intro_5} Despite these efforts, prolonged service disruptions, including withdrawal restrictions, made continued operations unsustainable.\cite{intro_5} Consequently, in December 2024, the company officially announced the cessation of its exchange operations and initiated the transfer of customer accounts and assets to SBI VC Trade, an exchange under Japan’s SBI Group. This case served as a sobering example that a single hacking incident could directly lead to the downfall of a business and the erosion of customer trust, underscoring the critical importance of security.

\subsubsection{2022 FTX Hack}

Founded in 2019, the globally renowned cryptocurrency exchange FTX collapsed in November 2022 due to financial mismanagement and executive misconduct, filing for bankruptcy.\cite{intro_1} In the immediate aftermath, a significant hacking incident occurred, exploiting internal chaos.\cite{intro_4} Approximately USD 477 million in various cryptocurrencies was illicitly transferred out of FTX wallets. Blockchain transaction analyses indicated that a substantial portion of these funds were stolen through hacking.\cite{intro_4} Experts suggested that the attackers had deep access to FTX’s internal security systems, evidenced by the use of verified accounts on another exchange, Kraken, to move funds. This led to speculation that the hack was likely perpetrated by insiders or through the abuse of privileged access.\cite{intro_1,intro_4} FTX swiftly transferred remaining assets to offline cold wallets and issued emergency advisories urging users not to access the FTX site or app.\cite{intro_4} The new management team coordinated closely with global law enforcement and other exchanges to track the stolen funds and mitigate further losses.\cite{intro_4} The FTX hack demonstrated that even large exchanges were vulnerable to massive customer asset losses due to poor internal controls and weak key management, further eroding trust amid the firm’s bankruptcy proceedings.

\subsubsection{2018 Coincheck Hack}

In January 2018, Japan’s major cryptocurrency exchange Coincheck was hacked, resulting in the theft of approximately USD 534 million worth of NEM (XEM) tokens—the largest known cryptocurrency theft at that time.\cite{intro_1} Coincheck immediately suspended all cryptocurrency deposits and withdrawals as it assessed the damage.\cite{intro_1} The attackers employed sophisticated phishing techniques to hijack employee accounts, planting malware into internal systems.\cite{intro_1} This enabled them to drain significant amounts of cryptocurrency from internet-connected hot wallets.\cite{intro_3} The exchange stated that it could not guarantee full compensation for customer losses, prompting an urgent investigation by Japan’s Financial Services Agency. At the time, some security experts and regulatory bodies suspected involvement by North Korea-linked hacking groups,\cite{intro_2} raising broader concerns about state-sponsored cyberattacks on financial platforms. The Coincheck hack highlighted the severe risks associated with inadequate hot wallet management and social engineering attacks, spurring stronger regulatory oversight and improved security practices within Japan’s cryptocurrency industry.

\subsubsection{2014 Mt. Gox Hack}

One of the earliest large-scale hacking incidents in cryptocurrency history occurred in early 2014, when the Tokyo-based exchange Mt. Gox lost approximately USD 460 million worth of Bitcoin.\cite{intro_1} At the time, Mt. Gox handled nearly 80\% of global Bitcoin trading volume, making it the largest exchange worldwide. However, repeated security breaches and operational mismanagement led to cumulative losses, culminating in a bankruptcy filing in 2014 that left around 24,000 customers without their deposited funds.\cite{intro_1} Post-incident investigations revealed that a lack of version control in the platform’s source code and other systemic vulnerabilities facilitated the attackers’ prolonged infiltration.\cite{intro_2} The case was widely regarded as the first major instance of a cryptocurrency exchange collapse driven by a hacking incident, and it had a lasting impact on the development of stronger security regulations and practices across the industry.


\subsection{Motivation}

\hspace*{1em}Despite more than a decade of academic advances and industry best-practices, private-key theft remains the dominant root cause of cryptocurrency losses. Chainalysis reports that hackers stole roughly USD 2.2 billion worth of digital assets in 2024 alone, marking a 21 \% year-over-year increase and the fifth year in the past decade to breach the billion-dollar threshold.
\cite{chainalysis2024stolen}
 Detailed incident analysis further reveals that compromised private keys account for 43.8 \% of all stolen funds, outranking smart-contract bugs and oracle manipulation.
\cite{chainalysis2025crime}
 The prevalence of this vector underscores a systemic weakness in the way keys are generated, stored, and used across both centralized exchanges and self-custody wallets.

Centralized services remain a lucrative target, with high-profile breaches such as the USD 245 million wallet drain that culminated in a kidnapping plot against the perpetrator demonstrating the real-world stakes of private-key leakage.
\cite{newstimes2025chetal}
 Even “cold” infrastructures can be circumvented: the February 2025 Bybit incident showed that multi-signature cold wallets are still vulnerable via supply-chain attacks on the hardware used during key generation.
\cite{cointelegraph2024multisig}
Self-custody is not inherently safer. Hardware wallets like Ledger or Trezor protect keys at rest, but phishing campaigns and firmware backdoors continue to siphon seeds once a device is connected to an online host.
\cite{cointelegraph2024hacks}
 Attempts to externalize keys—e.g., storing them on air-gapped USB drives or physical vaults—ultimately fail to remove the “last-mile” risk: the key must be decrypted inside a general-purpose CPU when signing a transaction, momentarily exposing it to rootkits, DMA attacks, or side-channel leakage.

TEEs such as ARM TrustZone provide an attractive isolation boundary, and prior work has indeed deployed single-purpose wallets (e.g., Bitcoin-only or Ethereum-only) inside such enclaves. However, a recent survey of TEE deployments finds that these projects are monolithic, protocol-specific, and rarely interoperable.
\cite{paju2023soktee}
 In practice, a user who holds assets across ten distinct chains must install—and keep patched—ten independent wallet applications or devices, each with its own threat surface and update cadence.
This fragmentation introduces three explicit pain points:
\begin{enumerate}[label=\arabic*.]
  \item \textbf{Usability overhead} -- Users juggle multiple seed phrases, user interfaces, and firmware versions, increasing the likelihood of human error and social-engineering success.
  \item \textbf{Delayed support for emerging chains} -- Niche or rapidly evolving blockchains often lack the commercial incentive for vendors to implement dedicated TAs, leaving early adopters to fall back on insecure hot wallets.
  \item \textbf{Duplicated audit effort} -- Security analysts must separately vet each single-chain implementation, wasting resources and risking inconsistent assurance levels.
\end{enumerate}Taken together, the evidence indicates a two-fold gap in today’s key-management landscape:
\begin{enumerate}[label=\arabic*.]
  \item \textbf{Security Gap} – Neither centralized custody nor conventional self-custody reliably prevents private-key exposure during transaction signing, making large-scale thefts routine. Hardware isolation solutions reduce—but do not eliminate—the attack surface, and they remain vulnerable to sophisticated supply-chain or side-channel exploits.
  \item \textbf{Interoperability Gap} – Existing TEE-based wallets hard-code support for a single blockchain. This siloed architecture fails to accommodate the multi-asset portfolios common among modern users and leaves minor networks underserved.
\end{enumerate}
Consequently, there is a pressing need for a unified, extensible, and verifiably secure framework that (i) confines all key material within a hardware-backed trusted boundary across its entire life cycle and (ii) allows dynamic addition of new blockchain support without compromising the assurance of established modules. The remainder of this paper addresses precisely this gap by introducing a modular TEE-centric wallet platform—detailed in Section 3—that reconciles strong cryptographic isolation with user-driven flexibility.
\subsection{Our Approach}
\begin{figure*}[ht]
    \centering
    \includegraphics[width=0.85\linewidth]{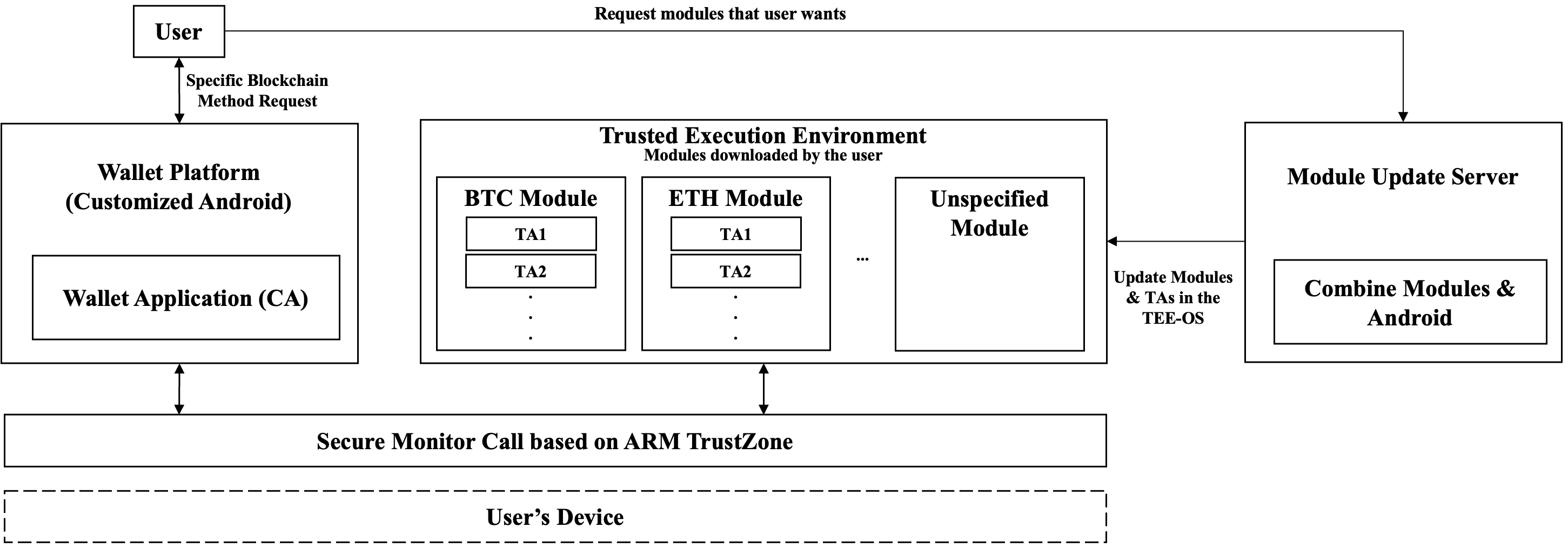}
    \caption{Overall solution of Modular Wallet}
    \label{fig:architecture}
\end{figure*}
\subsubsection{Design Goals}
The ultimate objective of this study is to provide a \textbf{key-management and signing infrastructure that is both secure and highly scalable in a multi-blockchain environment}, while simultaneously achieving user convenience and openness of the wallet ecosystem.  
To this end, we define the following four top-level goals.

\begin{enumerate}[label=\arabic*.]
  \item \textbf{Minimize key exposure}: Perform every secret-key operation inside the SoC’s ARM TrustZone–based Trusted Execution Environment (TEE) so that memory-snooping and code-injection vectors in both the application layer and the kernel space are fundamentally blocked.
  \item \textbf{Modularity}: Depart from the legacy paradigm that forces multiple protocols into a single monolithic Trusted Application (TA); instead, \textbf{package and distribute chain-specific TA modules independently}, thereby reducing functional dependencies and build complexity dramatically.
  \item \textbf{User-driven extensibility}: Users must be able to select and install only the network modules they need at any time, and even new or minor chains must be onboarded immediately with the same security level.
  \item \textbf{Verifiable supply-chain security}: Firmware-Over-The-Air (FOTA) images used for module distribution must undergo mandatory signature and integrity checks, and the loading procedure inside the TEE OS is standardized to prevent \textit{trust-on-first-use} attacks and post-installation tampering.
\end{enumerate}

\subsubsection{Architectural Overview}
Figure~1 illustrates the overall structure of the proposed system.  
The platform consists of three components:  
(i) a \textbf{Wallet Application} (Client Application, hereafter \textit{CA}) located in the Rich Execution Environment (REE) of the \textbf{user device};  
(ii) a set of \textbf{TA modules} residing in the Secure World of TrustZone; and  
(iii) a \textbf{Module Update Server}.
\begin{itemize}
  \item \textbf{Wallet Platform (Customized Android)}  
    \begin{itemize}
      \item Based on AOSP, only minimal syscall and HAL modifications are applied to expose an interface capable of invoking TrustZone Secure Monitor Calls (SMCs).
      \item The CA serializes user requests and forwards them to the TEE, then records the returned signatures either to the network broadcast layer or to local storage.
    \end{itemize}
  \item \textbf{Trusted Execution Environment}  
    \begin{itemize}
      \item Independent TAs for each chain are partitioned as the \textit{BTC Module}, \textit{ETH Module}, and so on.
      \item Each module encapsulates chain-specific logic such as key generation, address derivation, signing, and handling of BIP-32/44 parameters.
      \item Data sharing between TAs is prohibited to suppress side channels, and the public API is exposed only through the CA→TEE IPC interface.
    \end{itemize}
  \item \textbf{Module Update Server}  
    \begin{itemize}
      \item Registered developers submit signed module binaries and metadata (file hash, supported chain ID, interface version).
      \item The server runs static and dynamic analysis in a CI/CD pipeline to detect malicious code, hard-coded private keys, or excessive memory allocation.
      \item Upon passing all checks, a signed OTA package is generated and published to the version registry and CDN.
    \end{itemize}
\end{itemize}
\subsubsection{Module Lifecycle and Distribution Procedure}
\begin{enumerate}[label=\arabic*.]
  \item \textbf{Submission}: Developer X builds a module supporting \\“FooChain,” signs it with a GPG/PKCS~\#11 developer certificate, and uploads it to the server.
  \item \textbf{Validation}: The server automatically performs \textit{static analysis → symbolic execution → fuzzing}; packages in violation of policy are rejected.
  \item \textbf{Publishing}: Approved packages are deployed to the\\\texttt{modules.kms.example.com} repository, and the Manifest JSON is updated.
  \item \textbf{Installation}: When the user checks the FooChain option in the Wallet UI, the CA pulls down the OTA and streams the package into the Secure World.
  \item \textbf{Loading}: The TEE OS verifies the RSA-PSS signature and SHA-256 hash, then maps and executes the TA in an independent address space.
  \item \textbf{Update}: When a new version is posted, the CA downloads a delta-OTA in the background and hot-swaps it after the same verification procedure.
\end{enumerate}
\subsubsection{Threat Model and Countermeasures}
\begin{center}  
  \centering
  \footnotesize
  \setlength{\tabcolsep}{3pt}
  \renewcommand{\arraystretch}{1.15}

  \begin{tabularx}{\columnwidth}{|>{\raggedright\arraybackslash}p{2.8cm}|X|X|}
    \hline
    \textbf{Attack Vector} &
    \textbf{Legacy Monolithic Wallet} &
    \textbf{Proposed Mitigation} \\
    \hline
    REE Malware (sniffing) &
    Key operations exposed $\rightarrow$ theft possible &
    Perform all key operations inside the TA \\
    \hline
    OTA Supply-chain &
    Malicious update injection possible &
    Two-stage signatures (developer $\rightarrow$ server), TEE-OS-level verification \\
    \hline
    Side-channel (between TAs) &
    Shared process $\rightarrow$ leakage &
    TA isolation; L1/L2 cache partitioning; SG(PAN) enabled \\
    \hline
    Delayed support for new chains &
    Wallet vendor patch required &
    Community-driven third-party module submission and verification \\
    \hline
  \end{tabularx}
  \captionof{table}{Comparison of attack vectors, legacy limitations, and proposed mitigations.}
  \label{tab:attack_vectors}
\end{center}

\subsubsection{User Experience and Operational Perspective}
\begin{itemize}
  \item \textbf{Single UI, Multi-Chain}: Users manage \textit{BTC}, \textit{ETH}, \textit{Solana}, and \textit{FooChain} wallets simultaneously within one app, while the keys are guaranteed by the TEE.
  \item \textbf{Selective Installation}: Chains used infrequently can be disabled to reduce storage and memory consumption.
  \item \textbf{Backward Compatibility}: When importing legacy wallets, the CA feeds WIF/BIP-39 seeds into the TEE via a secure channel and immediately zero-izes them in memory.
  \item \textbf{Minimal Operator Liability}: The server performs only module verification and metadata distribution; private keys are never uploaded to the server.
\end{itemize}

\subsubsection{Technical Contributions}

\begin{enumerate}[label=\arabic*.]
  \item \textbf{Modular TEE stack design}: Overcame the \textit{single-binary TA} limitation of OP-TEE and implemented a \textit{multi-tenant TA store}.
  \item \textbf{Secure OTA pipeline}: Combined TA images and Android packages into a \textbf{single FOTA image}, unifying the user upgrade procedure.
  \item \textbf{Community-driven extensibility}: Demonstrated that even minor chains can be adopted immediately under the same security guarantees via a “verified contribution” model.
\end{enumerate}

\subsubsection{Summary}
This paper re-architects the key-management weakness of wallets into a \textbf{TEE-based modular architecture} that  
(i) fundamentally blocks key-exposure attack surfaces,  
(ii) provides multi-chain scalability and user-driven update flows, and  
(iii) simultaneously mitigates supply-chain threats and legacy compatibility issues.  
Our approach enables a new paradigm—\textit{“wallet functionality as a platform-level service”}—and experimentally proves that it can achieve both interoperability and security at scale in the broader blockchain ecosystem.

\section{Background}
\subsection{Blockchain}
\begin{figure}[h]
  \centering
  \includegraphics[width=\linewidth]{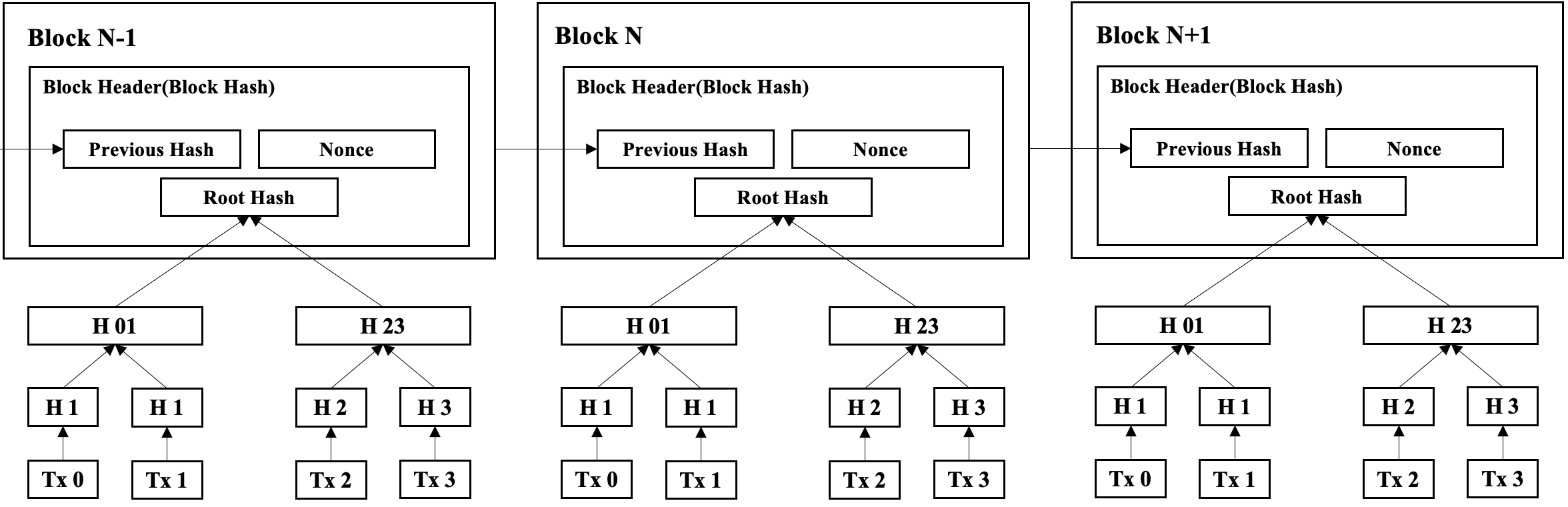}
  \caption{Structure of Blockchain.}
\end{figure}
\begin{table*}[ht]
\centering
\begin{tabularx}{\textwidth}{|c|c|X|c|X|}
\hline
\textbf{Generation} & \textbf{Period} & \textbf{Core Concept} & \textbf{Representative Projects} & \textbf{Technological Features and Development Direction} \\
\hline
1st Generation & Since 2009 & Decentralized Digital Currency & \textbf{Bitcoin} &
\begin{itemize}[leftmargin=*, nosep]
    \item Consensus algorithm based on PoW
    \item Ensures immutability and integrity of transactions
    \item P2P transfer system without central authority
\end{itemize} \\
\hline
2nd Generation & Since 2015 & Smart Contracts and DApps & \textbf{Ethereum} &
\begin{itemize}[leftmargin=*, nosep]
    \item Smart contracts using Turing-complete languages
    \item Decentralized applications (DApps)
    \item Token ecosystem (e.g., ERC-20)
\end{itemize} \\
\hline
3rd Generation & Since 2018 & Scalability, Interoperability, Sustainability & \textbf{\makecell[l]{Polkadot, Cardano, Solana, etc.}} &
\begin{itemize}[leftmargin=*, nosep]
    \item New consensus structures such as sharding and DAG
    \item Adoption of cross-chain technology
    \item Transition to PoS for energy efficiency
\end{itemize} \\
\hline
\end{tabularx}
\caption{\textbf{Blockchain Generational Classification}}
\end{table*}
\subsubsection{Concept of Blockchain} 
A blockchain is a form of distributed ledger technology that stores transaction records in data bundles called blocks, which are linked together in a continuous chain\cite{tripathi2023comprehensive}\cite{zheng2017overview}. Each block is cryptographically connected to the previous block by including the previous block’s hash value, thereby ensuring a strong bond and data integrity across the chain\cite{nakamoto2008bitcoin,tripathi2023comprehensive}. When new transactions occur, they are broadcast to the network, collected into a new block, and then appended to the existing chain in chronological order\cite{tripathi2023comprehensive}. All participating nodes employ a consensus algorithm to agree on which candidate block should be added to the blockchain, and once a block is finalized and added, the transactions it contains become extremely difficult to alter or remove. This design makes the blockchain an append-only ledger that is virtually tamper-proof, since any attempt to modify a recorded block would require redoing the proof-of-work for that block and every subsequent block\cite{nakamoto2008bitcoin,zheng2017overview}. Moreover, a blockchain operates as a decentralized network without a central administrator: the validity of transactions and updates to the ledger are determined collectively by network participants through cryptographic protocols\cite{zheng2017overview}. This allows parties to transact and share data in a trustless manner, meaning participants need not trust each other or any central entity – trust is instead placed in the blockchain’s cryptographic rules and decentralized consensus mechanisms\cite{zheng2017overview}.
\subsubsection{History of Blockchain}
The concept of blockchain was first introduced in 2008 when Satoshi Nakamoto published the Bitcoin white paper, which outlined a new peer-to-peer electronic cash system\cite{nakamoto2008bitcoin}. Nakamoto described an innovative method of timestamping transactions by hashing them into an ongoing chain of proof-of-work, thus forming a record that cannot be changed without redoing the proof-of-work\cite{nakamoto2008bitcoin}. In 2009, the Bitcoin network went live as the world’s first public blockchain, demonstrating that it was possible for participants to transfer digital currency in a P2P fashion without relying on any central authority\cite{tripathi2023comprehensive}. Bitcoin (now considered the first generation of blockchain technology) proved the viability of decentralized digital money and laid the foundation for blockchain development. In 2015, Ethereum launched, introducing the concept of smart contracts and marking the second generation of blockchain platforms\cite{tripathi2023comprehensive}. With smart contracts, blockchains could support programmable transactions and decentralized applications (DApps) beyond simple currency transfers, greatly expanding the use cases of the technology. Subsequently, newer projects focusing on scalability, interoperability, and other improvements emerged, often dubbed the third generation of blockchains. These developments extended blockchain applications into domains like supply chain, healthcare, and government services. In less than a decade, blockchain technology rapidly evolved through multiple generations, and research continues on enhancing performance, privacy, and sustainability of blockchains. Today, blockchain has grown from Bitcoin’s niche system into a broad foundational technology adopted or explored in various industries worldwide\cite{tripathi2023comprehensive}.

\subsubsection{Technical Components of Blockchain} 
A blockchain is built from several key technical components working in unison. The fundamental unit is the block, which groups a set of transactions along with metadata about those transactions\cite{tripathi2023comprehensive}. Typically, a block consists of a block header and a list of transaction contents. The block header contains information such as the hash of the previous block, a timestamp, a nonce or difficulty indicator (for proof-of-work), and the Merkle root which summarizes all transactions in the block\cite{tripathi2023comprehensive}. Notably, the inclusion of the previous block’s hash in each header links blocks together and creates the chained structure of the blockchain, whereby each block “points” to its predecessor. This hash chaining is critical for security: if any detail in an earlier block is altered, its hash changes and breaks the chain linkage, alerting the network to tampering.
Another core component is the transaction. A transaction represents a state change or transfer of value initiated by a user. For example, in a cryptocurrency blockchain, a transaction might transfer coins from one address to another. Each transaction generally contains the sender’s address, the recipient’s address, the amount (or data payload), and the sender’s digital signature, which proves the transaction was authorized by the holder of the corresponding private key. New transactions are broadcast to the entire network and collected into a pool of pending transactions (often called the mempool) by nodes. Miners or validators then select pending transactions to form a new block, which they propose to add to the blockchain\cite{tripathi2023comprehensive}.
The consensus mechanism is the protocol that network participants use to agree on the single authoritative ledger state in a distributed, trustless environment\cite{zheng2017overview}. In a public blockchain where participants may be anonymous or malicious, consensus algorithms ensure that all honest nodes converge on the same valid block to append next. The most well-known consensus algorithm is Proof of Work (PoW), used by Bitcoin, which requires nodes (miners) to perform a computationally intensive puzzle (hash calculation) to earn the right to create a block\cite{tripathi2023comprehensive}. PoW relies on the principle that finding a solution is difficult but verifying it is easy; the first miner to find a valid hash meeting the network’s difficulty target can broadcast their block, and other nodes quickly verify the proof-of-work and accept the block if it’s valid\cite{tripathi2023comprehensive}. PoW offers strong security and decentralization but at the cost of high computation and energy consumption. To address these costs, Proof of Stake (PoS) was developed and adopted by platforms like Ethereum, replacing brute-force computation with a system where validators stake cryptocurrency to win block creation rights\cite{tripathi2023comprehensive}. In PoS, validators are pseudo-randomly chosen to propose or validate blocks based on their stake (and other factors), and honest behavior is enforced by economic incentives (rewarding correct validation and slashing the stake for malicious behavior)\cite{tripathi2023comprehensive}. Numerous other consensus algorithms exist: Delegated Proof of Stake (DPoS), where stakeholders elect delegates to produce blocks; Practical Byzantine Fault Tolerance (PBFT), designed for consortium blockchains to tolerate a limited number of Byzantine (malicious) nodes; Proof of Activity, Proof of Burn, Proof of Capacity, and hybrid models, each with trade-offs in terms of speed, security, and decentralization\cite{tripathi2023comprehensive}. The choice of consensus mechanism is crucial and is often tailored to the blockchain’s use case – for example, PoW for permissionless security, or PBFT for permissioned efficiency. Together, the block data structure, transactions with cryptographic signatures, and the consensus protocol form the technical backbone of a blockchain system.

\subsubsection{Security Properties of Blockchain} 
Blockchain is lauded for its security and trust characteristics, which stem from its unique design principles and cryptographic foundations. First, a blockchain is decentralized, meaning no single entity controls the ledger; instead, control is distributed across many independent nodes that jointly validate and record data\cite{zheng2017overview}. This decentralization provides robustness: there is no single point of failure, and the network can continue to operate correctly even if some nodes are compromised or offline. Second, blockchains offer immutability – once data is recorded and confirmed on the chain, it is extremely difficult to alter. Each block is locked in place by the cryptographic hash linking it to the next block, so any attempt to change a past block would require recomputing the hashes of that block and all subsequent blocks on a majority of nodes, which is computationally infeasible in a large network\cite{nakamoto2008bitcoin}. This ensures that transaction records, once finalized, become tamper-evident and tamper-resistant, providing a permanent audit trail\cite{nakamoto2008bitcoin}. Third, blockchains are transparent and verifiable. In public blockchains, the ledger is openly available for anyone to inspect, and every node can independently verify the validity of transactions and blocks using the prescribed cryptographic checks[4]. 

\hspace*{1em}This openness promotes trust through verification – participants do not need to trust a central party, but can trust the system’s transparency and mathematics[4]. Fourth, blockchain users benefit from a degree of anonymity or pseudonymity. Identities on the blockchain are represented by addresses derived from public keys rather than personal information, so users can transact without revealing their real-world identity, yet every transaction is still linkable to a pseudonymous address[4]. This provides privacy for users to an extent, while still maintaining accountability since all transactions are traceable on the ledger (address identities can be audited even if the individuals behind them are not immediately known)\cite{zheng2017overview}. Lastly, blockchain security relies heavily on cryptographic techniques for integrity and authentication. All transactions must be digitally signed by the owner’s private key, which means that only the holder of the corresponding private key can authorize a given action (e.g., spending funds)[4]. Any alteration to a transaction’s data will invalidate its digital signature, allowing nodes to detect tampering instantly. This cryptographic validation, combined with hash linking of blocks and decentralized consensus, eliminates the need to trust intermediaries and instead places trust in robust algorithms. In summary, the key security properties of blockchain include decentralization (no single authority control), immutability of records, transparency and auditability, user pseudonymity, and cryptographic integrity assurance, all of which contribute to blockchain’s reputation as a trustworthy and secure data management solution\cite{zheng2017overview}.
\subsection{Signature}
\begin{table*}[ht]
\centering
\renewcommand{\arraystretch}{1.4}
\begin{tabular}{|l|p{3.5cm}|p{4.2cm}|p{6.5cm}|}
\hline
\textbf{Network} & \textbf{Key Curve / Size} & \textbf{Signature Algorithm} & \textbf{Address Format} \\
\hline
Bitcoin & secp256k1 (256-bit) & ECDSA on secp256k1 & Base58Check (P2PKH): \newline RIPEMD160(SHA256(pubkey)) ($\sim$20B, “1...” prefix)  \\
\hline
Ethereum & secp256k1 (256-bit) & ECDSA on secp256k1 & Hex (Keccak256(pubkey) low 20 bytes, “0x...” prefix) \\
\hline
Algorand & Curve25519 / Ed25519 \newline (256-bit) & EdDSA (Ed25519) \newline  & Base32 (public key + 4B checksum) \newline (58-char string)  \\
\hline
Polkadot & Curve25519 (256-bit) & Schnorrkel / Sr25519 \newline (Schnorr on Ristretto) & SS58 (Base58 of public key with prefix and checksum) ($\sim$48-char) \\
\hline
\end{tabular}
\caption{Comparison of Blockchain Key and Signature Schemes}
\label{tab:blockchain_signature_comparison}
\end{table*}
\subsubsection{Bitcoin (secp256k1 / ECDSA)}
Bitcoin uses the secp256k1 elliptic curve (defined by $y^2=x^3+7$ over a finite field) and the ECDSA signing scheme. A 256-bit private key $d$ is chosen uniformly at random, and the 65-byte (uncompressed) public key $P = dG$ is computed by scalar multiplication of the generator $G$. To sign a message hash $m$, one selects a random nonce $k\in[1,n-1]$, computes $R = kG$, then sets
\[
r = (x_R \bmod n), \quad
s = k^{-1}\!\bigl(H(m) + d\,r\bigr) \bmod n,
\]
so the signature is $(r,s)$. The security relies on the Elliptic Curve Discrete Logarithm Problem. Bitcoin addresses are derived by hashing the public key: for a standard Pay-to-Public-Key-Hash (P2PKH) address, one computes $\mathit{HASH160}(P)=\mathrm{RIPEMD160}(\mathrm{SHA256}(P))$ and encodes it in Base58Check with the network prefix. The result is a 160-bit address (typically 26–35 characters, starting with “1” for mainnet) that includes a 4-byte checksum. 
\subsubsection{Ethereum (secp256k1 / ECDSA)}
Ethereum also uses secp256k1 and ECDSA (with the same signing math as Bitcoin).A 256-bit private key yields a point $P=dG$ on the curve $y^2=x^3+7$. To generate an account address, the 64-byte uncompressed public key is hashed with Keccak-256, and the low-order 20 bytes (160 bits) of the hash are taken as the address. Formally, $\mathit{Addr} = \mathrm{Keccak256}(P)$, yielding the familiar 20-byte hex address (prefixed “0x”). The ECDSA algorithm allows public-key recovery from the signature (using a recovery id $v$), which is why Ethereum stores $(r,s,v)$. In summary, Ethereum’s address generation and signing mirror Bitcoin’s curve and signature (ECDSA on secp256k1) but use Keccak-256 hashing instead of RIPEMD-160.
\subsubsection{Algorand (Ed25519 / EdDSA)} Algorand uses the Ed25519 signature scheme (an instance of EdDSA) over the Edwards-form Curve25519. The private key $a$ and its public key $A=aB$ (where $B$ is a curve base point) are 32-byte values. The Ed25519 signing procedure (as described by Bernstein et al.) is: compute $r = H(k \parallel M)$ from the private key’s prefix and message $M$, then $R = rB$ and $S = (r + H(R,A,M),a)\bmod \ell$, yielding signature $(R,S)$. This yields $\approx 256$-bit security. Algorand addresses are directly derived from the public key: one appends a 4-byte checksum to the 32-byte Ed25519 public key and encodes the result in Base32. Thus an Algorand address is essentially the public key plus checksum, typically represented as a 58-character Base32 string.
\subsubsection{Polkadot (Sr25519 / Schnorr)}. Polkadot (and Substrate chains) support multiple signature schemes, but their default is Schnorrkel (often called sr25519). This is a Schnorr-type signature on the same Curve25519 used by Ed25519. In sr25519, a private key $d$ yields a public point $P=dG$, and signing is similar to Schnorr: pick random $k$, compute $R=kG$, and set $s = k + H(R,P,m),d\pmod q$. (Polkadot’s implementation includes specific tweaks for cofactor and hashing, but conceptually it is Schnorr.) The advantage is native multi-signature support and efficiency. Addresses in Polkadot use the SS58 format: the public key (32 bytes) is prefixed by a version byte and checksum, then Base58 encoded. For example, Polkadot mainnet addresses start with a specific prefix (usually capital “1” or “D” depending on version) and are ~48 characters long. (Ed25519 and ECDSA keys are also supported by Polkadot, but sr25519 is the recommended scheme.)

\subsection{Trusted Execution Environment(TEE)}

\hspace*{1em}A Trusted Execution Environment (TEE) is a hardware-assisted secure area of a processor that guarantees code and data loaded inside it are protected with respect to confidentiality and integrity. The primary goal of a TEE is to allow sensitive computations to execute in isolation, even if the host operating system or hypervisor is compromised. In practice, a TEE creates a programming environment that keeps critical assets (code, keys, data) safe from certain attacks which would be difficult to thwart with traditional software security alone\cite{costan2016sgx}. This is achieved by partitioning hardware and software resources into protected and unprotected domains, with hardware-enforced access control between them\cite{costan2016sgx}. In essence, the TEE forms a “secure world” that normal (untrusted) software cannot directly access, establishing a strong security perimeter.

TEEs are designed with isolation mechanisms at their core. These mechanisms ensure that any code running outside the TEE cannot inspect or tamper with the code and data inside the TEE. This isolation can be implemented via dedicated hardware privileges, memory access controls, and often memory encryption. For example, Intel’s Software Guard Extensions (SGX) technology creates protected containers called enclaves in a process’s address space; these enclaves are opaque to the rest of the system and remain confidential and integer even in the presence of a malicious OS\cite{mckeen2013hasp}. In general, the TEE’s hardware will block or encrypt unauthorized accesses, so that even highly privileged malware (kernel or hypervisor level) cannot read or modify TEE memory. Many TEEs also support secure attestation mechanisms: the TEE can produce cryptographic proof of what software is running inside it. This attestation allows a remote party to verify the TEE’s contents and establish trust before provisioning sensitive data. For instance, an application running in an SGX enclave can prove its identity (enclave code hash) to a remote server and then receive decryption keys or credentials over an encrypted channel\cite{mckeen2013hasp}. In summary, TEEs provide a foundation for secure computation on untrusted platforms – sometimes termed confidential computing – by combining hardware-based isolation, attestation, and minimal trusted code bases.

\subsubsection{Intel Software Guard Extensions (SGX)}

\hspace*{1em}Intel SGX is a prominent TEE technology introduced by Intel for x86 processors (first available in 2015). SGX extends the CPU instruction set to allow the creation of protected memory regions called enclaves within user-level applications. An enclave is a reserved portion of a process’s address space that is protected by the CPU such that even the OS, hypervisor, or BIOS cannot read or alter its contents\cite{mckeen2013hasp}. Any attempt by external software to access enclave memory is blocked by hardware, ensuring that enclave code and data enjoy confidentiality and integrity guarantees even if privileged software is malicious. The enclave memory is encrypted in DRAM by a Memory Encryption Engine and checked for integrity (e.g., via Merkle-tree mechanisms), so that hardware ensures enclave memory can’t be spoofed or read in plaintext outside the CPU package. Notably, enclaves are created and managed by the untrusted OS (which allocates pages and schedules threads), but the OS cannot violate enclave protections – this design minimizes the Trusted Computing Base (TCB) to include only the enclave’s own code and the CPU/hardware. By dramatically reducing the TCB, SGX limits the attack surface: a compromised operating system may deny service but cannot extract secrets from an enclave or alter its execution results.

SGX enclaves are primarily intended to enable secure cloud and application scenarios where the developer partitions an application into security-critical components that run inside enclaves. Communication across the enclave boundary is done via well-defined calls; if an enclave needs to request OS services or perform I/O, it must exit, which ensures the enclave’s internal state is not directly exposed. To bolster trust, SGX supports a remote attestation service: each enclave can produce a quote (digital signature) over its measurement (a cryptographic hash of its initial code/data) using a processor-resident key. Remote clients verify this quote (via Intel’s attestation infrastructure) to confirm they are communicating with a genuine Intel SGX enclave running the expected code, then provision secrets (e.g. decryption keys) to it\cite{mckeen2013hasp}. Once provisioned, the enclave can use those secrets internally, and even “seal” data to disk encrypted with a hardware-bound key such that only the same enclave (or enclave author) can unseal it later. In terms of performance, SGX enclaves incur some overhead (for enclave transitions and memory encryption) and are limited in secure memory size (the Enclave Page Cache is on the order of tens of MBs in early SGX versions). Nevertheless, SGX provides a powerful level of isolation: even a kernel-level attacker or a malicious cloud administrator cannot directly read enclave memory contents \cite{mckeen2013hasp}. This strong protection has made SGX a key example of fine-grained application-level TEEs focused on high security for sensitive code.

\subsubsection{ARM TrustZone}

\hspace*{1em}ARM TrustZone takes a different approach to TEEs, providing a coarse-grained division of the entire system into two execution worlds. Introduced in ARM processors (initially for ARMv6 architecture), TrustZone hardware creates a Secure world and a Normal world that run in parallel on the same processor core. The Normal world hosts the rich operating system (e.g., Android or Linux) and regular applications, while the Secure world runs a small, trusted OS (often called a Trusted Execution Environment OS) and secure applications (trustlets). The transition between worlds is controlled by the processor: a special Secure Monitor mode mediates switches typically via a secure monitor call (SMC) instruction. Critically, the hardware enforces that Normal world code cannot access Secure world memory or devices: all physical memory and peripherals are tagged as Secure or Non-secure, and the TrustZone address controller will reject or remap accesses from the Normal world to secure resources\cite{arm2009trustzone}. This means the Secure world remains isolated from the rich OS, creating a safe haven for security-sensitive code (like cryptographic key managers, DRM agents, payment modules, etc).

The TrustZone model essentially provides a privileged isolated environment for an entire small operating system. In practice, a typical TrustZone Secure world runs a minimal Trusted OS (such as ARM’s Trusted Firmware or proprietary solutions like Qualcomm’s TrustZone TEE OS), and that OS can host multiple trusted applications. All Normal world software is untrusted and cannot directly interfere with the Secure world; any interaction (for example, a banking app requesting a fingerprint authentication handled in secure world) must be done via calls mediated by the secure monitor. Because the Secure world operates at a higher privilege (secure supervisor mode) than the Normal world OS, it can also observe or vet certain normal world operations if needed. The TEE provided by TrustZone thus allows device manufacturers and OS vendors to implement a range of secure services (secure UI, biometric processing, digital rights management, etc.) that are protected from potentially compromised normal applications or kernels\cite{arm2009trustzone}. One advantage of TrustZone is its low performance overhead: since the secure and normal environments time-share the same core (with context switches on world transitions) and use the same instruction set, switching worlds is relatively fast and there is no need for duplicate hardware. Also, existing applications in the Normal world run at full speed, unaware of TrustZone’s presence unless they invoke secure services.

However, TrustZone’s approach has some trade-offs. The granularity of isolation is system-wide – there is only one Secure world, so all trusted components share the Secure world’s resources. This means the TCB in TrustZone includes the secure kernel and all secure applications (which could be sizable, potentially millions of lines of code when a rich TEE OS is used). A vulnerability in the Secure world could compromise all secure apps. In contrast, SGX enclaves are isolated from one another at the hardware level; TrustZone must rely on its secure OS to compartmentalize different secure apps. Nonetheless, TrustZone remains very popular in mobile and embedded devices due to its ease of integration: it does not require major changes to normal-world apps or OS (aside from using the API to call secure services), and it leverages the existing processor with minimal extra silicon. It shines in scenarios like protecting device unlock mechanisms, payment credentials, or other localized secrets where the device manufacturer controls the secure world software. TrustZone can also be used in virtualization scenarios (e.g., a hypervisor in secure world and a guest OS in normal world), but it fundamentally assumes a single trusted realm and does not target multi-tenant cloud use cases the way SGX or SEV do.

\subsubsection{AMD Secure Encrypted Virtualization (SEV)}

\hspace*{1em}AMD SEV is a TEE technology from AMD that focuses on isolating entire virtual machines from a potentially untrusted hypervisor. Debuted with AMD’s EPYC processors (around 2017), SEV’s design aligns with cloud computing needs: it allows a guest VM’s memory to be encrypted such that the cloud provider’s hypervisor cannot read it. In AMD’s architecture, the memory of each VM is automatically encrypted by the CPU with a key unique to that VM. The hypervisor and host remain responsible for scheduling and managing VMs, but they only ever see ciphertext when accessing a guest’s memory. This creates a form of invisible VM to the hypervisor, enforced by hardware. Concretely, AMD’s memory controller tags each memory page with an identifier for the VM (ASID), and on each memory access it uses the tag to determine the appropriate key for encryption/decryption\cite{kaplan2021memoryencryption}. If a hypervisor or another VM tries to read memory not belonging to it, they will get only encrypted data. In this way, SEV provides strong cryptographic isolation between VMs and from the hypervisor\cite{kaplan2021memoryencryption}. The CPU and a secure co-processor (the AMD Secure Processor) manage the keys and ensure they remain inaccessible to software, so that even a rogue hypervisor with full control of the system cannot extract a guest VM’s memory in plaintext.

The strength of AMD SEV is that it can protect unmodified legacy applications and operating systems running inside virtual machines – the guest OS does not need to be rewritten or aware of SEV in most cases. For cloud customers, this means one can run their workload in a VM on someone else’s server and be confident that even the cloud provider’s admins or malware in the host cannot snoop on the VM’s memory contents. SEV technology evolved through several generations: the initial SEV provided memory encryption for guest pages, later SEV-ES extended protection to the VM’s CPU register state when the VM is switched out (encrypting CPU registers on VM exits so the hypervisor can’t capture them), and the latest SEV-SNP adds hardware-based memory integrity protection and other enhancements\cite{amd2020sev}. Memory integrity protection is crucial because pure encryption does not stop a malicious hypervisor from altering or replaying encrypted memory content. In fact, early SEV versions, while preventing the hypervisor from learning plaintext memory, could be vulnerable to tampering (e.g., replaying an old encrypted page into the VM). SEV-SNP addresses this by maintaining integrity checks so that any unauthorized modifications to encrypted memory are detected, preventing attacks like replay or memory re-mapping\cite{amd2020sev}. With SEV-SNP, AMD’s TEE approach comes closer in security strength to Intel’s enclaves, effectively creating an isolated execution environment for an entire VM with both confidentiality and integrity assured against the hypervisor.

The trust model in SEV differs from SGX. In SEV, the guest OS and applications are all inside the protected boundary (and are therefore part of the trusted domain from the perspective of an outside attacker). The hypervisor and any higher-level host software are untrusted. This means SEV does not protect an application from a malicious or compromised kernel within the same VM – if the guest OS is compromised, the applications in that VM are affected. In contrast, SGX aims to protect an enclave even from a malicious host OS. Thus, SEV’s threat model is suitable for scenarios where the infrastructure (hypervisor, cloud provider) is untrusted, but you still trust your own VM’s OS. AMD does provide a form of attestation as well: the AMD Secure Processor can generate evidence of a VM’s initial state (firmware, BIOS, etc., and measurement of VM image) which a cloud tenant can verify to ensure their VM was launched correctly on genuine AMD hardware before provisioning secrets. Overall, SEV is a coarse-grained TEE – its unit of isolation is an entire virtual machine – but it’s very practical for cloud deployment since it can secure existing software stacks with minimal performance impact (the encryption/decryption overhead on modern AMD CPUs is low, typically only a few percent performance cost). It complements the finer-grained TEEs like SGX by addressing a different use case: protecting the confidentiality of whole VMs in multi-tenant environments with untrusted hosts.

\subsubsection{Comparison of TEE Technologies (SGX vs. TrustZone vs. SEV)}

\hspace*{1em}Each of the above TEE technologies shares the common goal of isolating sensitive computations, but they differ in architecture and strengths:

\begin{itemize}[itemsep=0.25em, topsep=5pt, left=0pt]
    \item \textbf{Isolation Granularity and Scope:} Intel SGX provides process-level isolation, allowing many independent enclaves (even within one application) each protecting specific code and data. ARM TrustZone enforces a system-wide split into only two domains (secure vs normal), suitable for running an entire trusted subsystem alongside a normal OS. AMD SEV operates at the VM level, encapsulating an entire guest OS and its processes within a protected container (the VM). This means SGX can secure individual application modules, TrustZone secures platform services or drivers, and SEV secures whole OS instances.
    \item \textbf{Threat Model and TCB:} SGX assumes an attacker may control everything in the software stack (OS, hypervisor, BIOS) except the CPU; thus the enclave’s TCB is minimal (mainly the enclave code itself and CPU microcode)\cite{mckeen2013hasp}. TrustZone assumes the normal world OS may be compromised, but the secure world OS remains trusted – the TCB includes the secure OS and trusted apps, which can be relatively large. SEV assumes an untrusted hypervisor/host, but trusts the guest OS; the TCB for an application in an SEV-protected VM still includes the entire guest OS (and hypervisor firmware) even though the underlying hardware/firmware protects the VM from the hypervisor. In short, SGX offers the smallest TCB (no OS in TCB), TrustZone and SEV have a larger TCB (they include an OS inside the trusted region, secure OS for TrustZone, guest OS for SEV).
    \item \textbf{Isolation Mechanism:} TrustZone primarily uses logical separation – a CPU mode (secure world) and an NS bit for memory/peripherals – to enforce isolation\cite{arm2009trustzone}. It does not inherently encrypt secure world memory (it relies on the fact that normal world simply cannot access it at all). SGX uses a combination of CPU access control and memory encryption (with hardware-managed encrypted memory regions for enclaves) to protect enclave pages even if they reside in regular DRAM\cite{mckeen2013hasp}. AMD SEV relies on full memory encryption with per-VM keys to isolate VMs; only authorized hardware can decrypt a VM’s memory\cite{kaplan2021memoryencryption}. Initially SEV’s isolation was purely cryptographic (encrypting memory so the hypervisor can’t read it); with SEV-SNP it also includes integrity checks to prevent unauthorized modification\cite{kaplan2021memoryencryption}.
    \item \textbf{Attestation and Provisioning:} All three technologies provide ways to attest the secure environment. SGX has a well-defined remote attestation process for each enclave, allowing remote parties to verify enclave identity before provisioning secrets\cite{mckeen2013hasp}. TrustZone-based TEEs historically have had proprietary attestation (e.g., Samsung Knox or Google’s Titan M might attest certain secure world state), but there isn’t a universal remote attestation standard for TrustZone across all vendors. AMD SEV attestation involves the platform’s Secure Processor signing a statement of the VM’s launch state (covering the hypervisor, firmware, and VM image measurements) which the VM owner can verify – ensuring the VM is running under SEV protection on genuine AMD hardware. In summary, SGX and SEV enable cloud use-cases with remote attestation, whereas TrustZone’s attestation is typically used locally or within an ecosystem (e.g., phone manufacturer verifying the secure OS).
    \item \textbf{Performance and Use Cases:} TrustZone has virtually no runtime performance penalty for normal world code (aside from the overhead when switching into secure world), making it ideal for real-time or low-power device scenarios. Its use case is device-centric (protecting local secrets, secure user interface, etc.). SGX enclaves incur some performance cost (especially for I/O-intensive operations or frequent context switches) and have memory size limitations, but they enable powerful cloud applications like secure multi-party analytics or blockchain oracles by protecting small sensitive code chunks on any platform \cite{mckeen2013hasp}. AMD SEV has a slight performance overhead (memory encryption and maybe a small cost for cryptographic key use), but it can secure entire VMs with ease, which is highly valuable for cloud providers offering confidential VMs to customers. SEV is less useful if one does not use virtualization, whereas SGX/TrustZone can protect code in non-virtualized environments. Each TEE has strengths suited to different scenarios: SGX excels at fine-grained protection for specific application logic (with maximum trust in hardware), TrustZone excels at integrating security functions into devices with minimal disruption, and SEV excels at lifting and securing large legacy software stacks (whole operating systems) in cloud environments.
\end{itemize}

\subsection{Android Firmware Over-The-Air}

\subsubsection{FOTA Update System Architecture}

\hspace*{1em}Modern Android devices employ a robust Firmware Over-The-Air (FOTA) update system to upgrade the operating system and firmware without requiring physical access. The standard Android update process (as described in AOSP documentation) uses an A/B partition scheme for seamless updates\cite{aosp2025abupdate}. In essence, the device has two copies of critical partitions – usually referred to as slot A and slot B – such as system, boot, vendor, etc\cite{aosp2025abupdate}. At any given time, the device boots and runs from the active slot (say, slot A), while the other slot (slot B) is inactive/unused. During an OTA update, the new firmware is written to the inactive slot in the background, while the user can continue using the device normally on the active slot. This approach ensures that there is always a known-good system partition available; if anything goes wrong with the update, the device can fall back to the old slot, greatly reducing the risk of ending up with an unbootable device\cite{aosp2025abupdate}. This dual-partition design is often called seamless update because the installation can occur without taking the device offline (aside from a brief reboot), and it provides fail-safety.

\subsubsection{A/B Partition Scheme and Update Flow}

\hspace*{1em}In the standard update flow using A/B partitions, the process unfolds as follows:

\begin{enumerate}[itemsep=0.5em, topsep=5pt, left=0pt]
    \item \textbf{Download stage:} The device checks for an available update (usually via a software update client, such as Google Play Services on Android One/Pixel devices, or an OEM-specific updater). Once an update is available, the device downloads the OTA package (archive containing the update payload). The OTA package is typically a zip file with binary patches or full images for the target partitions. (Note: Android supports streaming OTAs, where the update can be applied as it downloads, to avoid needing large storage space for the whole package.)
    \item \textbf{Preparation stage:} Before applying the update, the system prepares the slots. The currently running slot (e.g. slot A) is marked as successful (if it wasn’t already) to ensure the bootloader will continue to trust it as a fallback\cite{aosp2025abupdate}. The inactive slot (slot B) is marked as unbootable initially\cite{aosp2025abupdate}, because its contents will be partially updated and thus inconsistent until the process is complete. This prevents the bootloader from accidentally booting the half-updated slot in case of a reboot during the update.
    \item \textbf{Installation stage:} The update daemon (Android’s update\_engine) writes the new update payload to the partitions of the inactive slot (B) without disturbing the active slot (A)\cite{aosp2025abupdate}. It performs a series of operations described by the update metadata, such as patching or replacing blocks of the B partitions with new data. Throughout this stage, the system on slot A keeps running. The update\_engine carefully coordinates reads from slot A and writes to slot B, and can pause/resume if the device reboots or loses power, since slot A is still intact. After all operations, update\_engine verifies the integrity of the updated slot B partitions by computing hashes and comparing with expected values from the metadata\cite{aosp2025abupdate}, ensuring the write was successful.
    \item \textbf{Switching stage:} Once the new firmware is fully written to slot B and verified, the device sets slot B as the active slot in the bootloader’s slot metadata (using the boot control HAL, i.e., calling setActiveBootSlot() routine)\cite{aosp2025abupdate}. However, at this point slot B is still not marked “successful” – it’s just ready to boot into. The system then initiates a reboot. (The user is typically prompted to reboot now to apply the update.)
    \item \textbf{Reboot and verification stage:} On reboot, the bootloader will load the new updated slot (B) as it has been marked active. Android’s Verified Boot process (based on dm-verity) then checks the cryptographic integrity of the slot B system, boot, and other partitions before handing control to the OS. If any corruption or tampering is detected (e.g., a bad flash), the boot will fail dm-verity verification. Notably, dm-verity guarantees the device only boots an uncorrupted image\cite{aosp2025abupdate}. If the new slot B fails to boot for any reason – whether a verification failure or a runtime crash – the bootloader will notice that the slot has not reported a successful boot and can automatically revert to slot A on the next reboot\cite{aosp2025abupdate}. (The bootloader uses metadata flags: it gives the new slot a “boot attempt” and if the slot doesn’t confirm success, it decreases a retry count. After a few failed tries, it marks the slot unbootable and falls back to the other slot.)
    \item \textbf{Post-boot stage:} If slot B boots correctly into the new Android version, a component in the system (often update\_verifier or the update client) will perform final checks and then mark the new slot B as successful by calling a bootloader interface (markBootSuccessful())\cite{aosp2025abupdate}. Marking it successful tells the bootloader that the slot is good and can be kept as permanent. At this point, the update is officially complete. The old slot A remains on disk as a backup until perhaps the next update, or the system may later use that storage for something else (some devices may retain the last known-good backup, others might eventually wipe an old slot when not needed).
\end{enumerate}

Throughout this flow, redundancy and verification are key: the active slot’s content is never modified during update (so the current system is safe if update fails), and all updated data is checksummed and cryptographically verified. The device also records the update progress so that if the process is interrupted (due to a power loss or user reboot), it can resume or roll back gracefully on the next boot. Users experience minimal downtime – the update installation happens in the background, and the only interruption is the reboot into the new firmware, which is roughly as fast as a normal reboot\cite{aosp2025abupdate}.


\section{Paper Review}
\subsection{OP-TEE Based TEE Framework: Open-TEE (McGillion et al., 2015)}

\hspace*{1em}Trusted Execution Environments (TEEs), leveraging hardware features such as ARM TrustZone and Intel SGX, provide a secure execution environment for protecting sensitive data.\cite{paper_1} They are increasingly used in blockchain wallets, digital certificate management, payment systems, and various other domains. 

However, despite their strong security guarantees, conventional hardware-based TEEs suffer from significant limitations in terms of development accessibility. To address these issues, McGillion et al. (2015) proposed *Open-TEE*, an open-source virtual TEE framework. The study implemented a hardware-independent, easy-to-develop and test TEE environment while strictly adhering to the GlobalPlatform (GP) standards. 

Traditionally, ARM TrustZone-based TEEs imposed high barriers to entry for developers and researchers due to licensing, development kit costs, and restricted access tied to chip vendors. McGillion et al. designed Open-TEE to lower these barriers, enabling anyone to develop and test TEE applications by running the platform entirely in Linux user space. In particular, Open-TEE allows developers to use familiar tools such as GCC, GDB, and OpenSSL, thereby significantly enhancing productivity in TEE application development. 

From an architectural standpoint, Open-TEE adopted an Android Zygote-like Launcher to optimize TA loading speed and implemented IPC mechanisms based on Unix Domain Sockets.
\begin{figure}[htbp]
    \centering
    \includegraphics[width=0.8\linewidth]{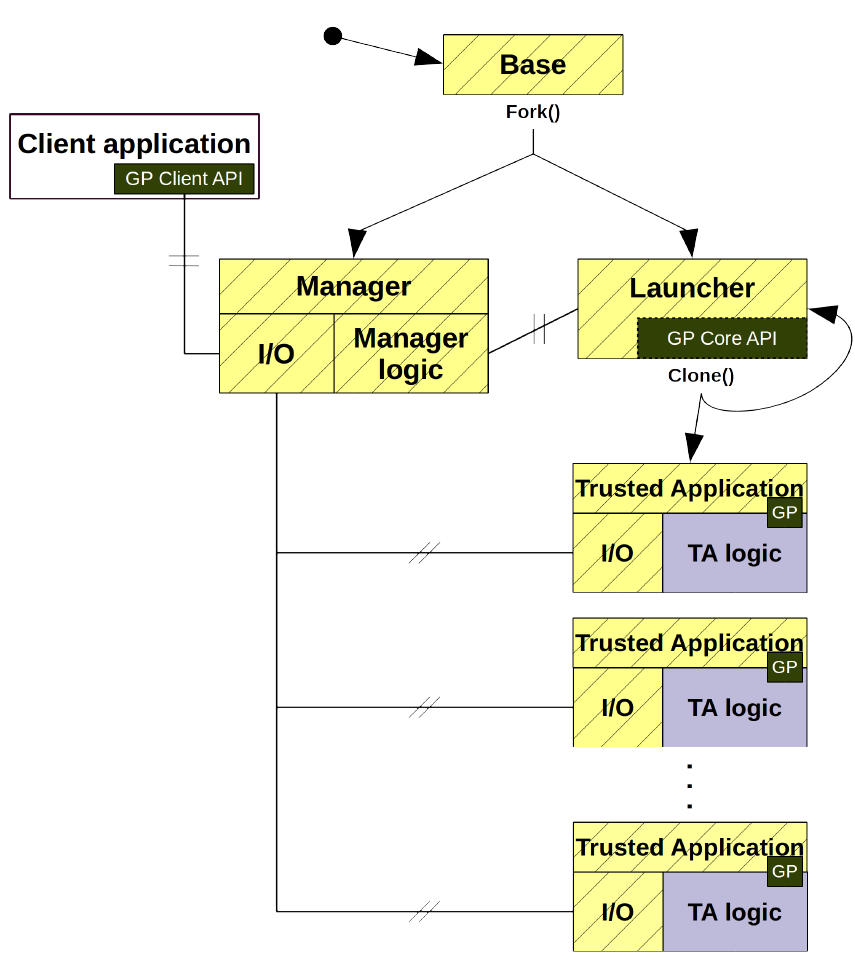}
    \caption{Open-TEE architecture}
    \label{fig1}
\end{figure}
This design provided a much faster build-execute feedback loop compared to conventional TEE systems, achieving an average build time of 147 ms and an execution time of 430 ~$\mu$s. Furthermore, while simulating the Secure Monitor Call (SMC) interface, Open-TEE provides GP-compliant Client and Core APIs to ensure structural compatibility when porting TAs to actual hardware-based TEEs. 

The study clearly distinguished Open-TEE from existing platforms such as QEMU-based TrustZone emulators, Trusted Little Kernel (TLK), and OP-TEE. Unlike OP-TEE and TLK, which require hardware-level porting at the device driver, Secure Monitor, and bootloader layers, Open-TEE operates entirely in user space, enabling rapid TEE application development without hardware dependencies. This allows developers to develop and debug TA code even without access to TrustZone-enabled hardware. 

Moreover, McGillion et al. conducted a user study to empirically validate the usability of Open-TEE. In a study involving 14 developers with TEE development experience, Open-TEE was shown to significantly improve usability compared to traditional hardware-based development environments. While issues such as frequent device resets and inefficient build-flash-execute cycles were commonly cited problems in existing environments, Open-TEE resolved these issues, improving both development speed and productivity. 

In the context of this research — which proposes a dynamic firmware update architecture that composes and applies TA modules on the server based on the user’s selected blockchain network — Open-TEE plays a key role. During the development of blockchain-specific TA modules, Open-TEE enables rapid prototyping and testing while maintaining GP-compliant architecture for seamless migration to OP-TEE or actual TrustZone-based hardware platforms. This offers an effective development methodology for verifying server-side TA composition processes, automating new firmware builds, and establishing pre-deployment verification environments. 

In conclusion, McGillion et al.’s Open-TEE represents a pioneering effort in improving TEE application development accessibility and serves as an essential technical foundation for realizing this study’s goal of a flexible, multi-blockchain, firmware update-based architecture.

\subsection{TrustZone-based SPV Wallet Design: SBLWT (Dai et al., 2018)}

\hspace*{1em}The most critical security requirements in blockchain wallet design are the secure storage and usage of private keys, and the assurance of integrity during transaction signing.\cite{paper_2} However, software wallets running in conventional mobile environments operate in the normal OS domain, making them vulnerable to OS-level malware, memory analysis, and debugging attacks. 

On the other hand, hardware wallets offer stronger physical security and isolation but suffer from usability issues due to the need for separate devices. To address these challenges, Dai et al. (2018) proposed *SBLWT (Secure Blockchain Lightweight Wallet based on TrustZone)* — a lightweight blockchain wallet architecture built on ARM TrustZone technology. 

The study specifically focused on integrating TrustZone-based TEE protections into SPV (Simplified Payment Verification) wallets, in order to enhance the security of private keys and transaction signing. While SPV-based mobile wallets reduce storage and network usage by maintaining only block headers rather than full blockchain data, the verification process itself typically runs in the normal OS domain, leaving it vulnerable to memory analysis and API hooking attacks. 

To overcome these limitations, Dai et al. migrated block header storage, key management, and transaction signing processes into the TrustZone-based Secure Execution Environment (SEE).
\begin{figure}[htbp]
    \centering
    \includegraphics[width=\linewidth]{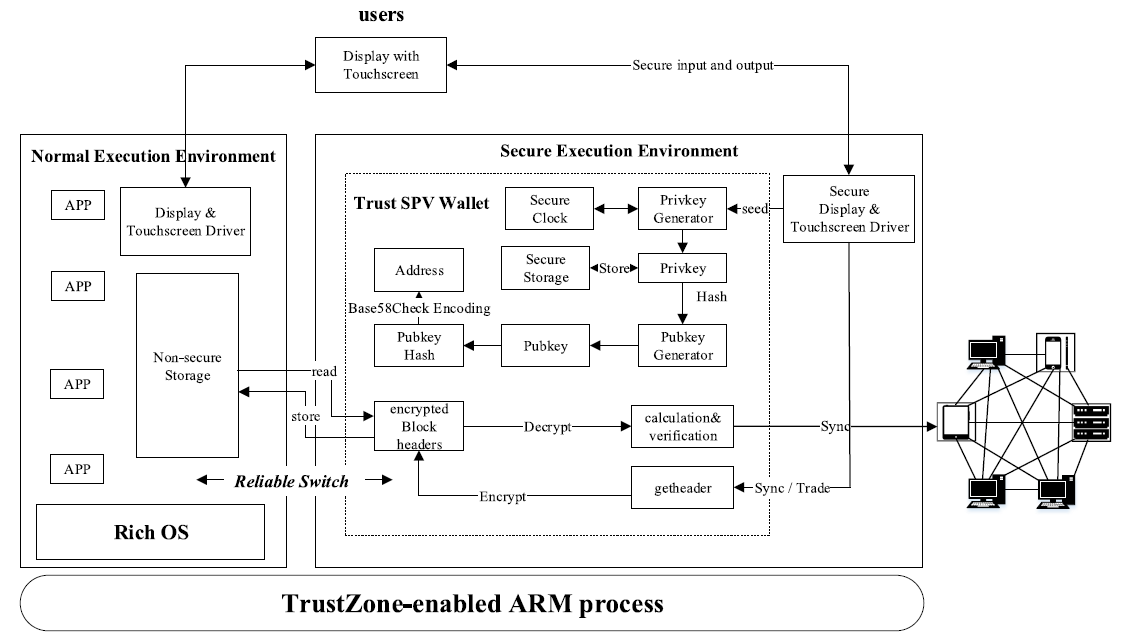}
    \caption{The framework of SBLWT.}
    \label{fig2}
\end{figure}

The SBLWT architecture is divided into two main domains:

\begin{itemize}[itemsep=0.25em, topsep=5pt, left=0pt]
    \item In the Secure World, private key generation and storage, block header management, transaction signing, and random number generation are performed.
    \item The Normal World handles UI interactions and network communications.
\end{itemize}

Block headers are stored in encrypted form in non-secure storage and can only be decrypted within the Secure World. Through this design, SBLWT eliminated the risk of private key exposure during blockchain verification and implemented protections against MITM (Man-in-the-Middle) attacks. 

In terms of secure UI, Dai et al. implemented TrustZone-based Secure Display and Secure Touchscreen drivers to prevent malicious apps in the normal OS from spoofing or tampering with transaction confirmation screens. This aspect, which had been relatively neglected in earlier mobile wallet research, greatly improved the security of user interactions. 

The system was implemented using OP-TEE and tested on a Raspberry Pi 3 Model B board. Performance evaluations demonstrated that running blockchain verification and transaction signing in the Secure World did not cause significant degradation and had negligible impact on the user experience. Furthermore, because no sensitive data was exposed outside the Secure World during SPV verification, the system offered significantly higher security than conventional software wallets. 

In their comparative analysis, Dai et al. noted that most prior TrustZone-based solutions had focused on full node wallets or cold wallets, with few examples of secure, lightweight SPV wallet implementations. Whereas commercial products such as Ledger and Rivetz relied on proprietary APIs or closed TEE implementations, SBLWT was built on the open-source OP-TEE platform and adhered to GlobalPlatform (GP) standard APIs, improving portability across platforms. 

In this study’s proposed architecture — which dynamically composes and updates TA modules on the server based on user-selected blockchain networks — SBLWT offers important insights. 

When designing TA modules for different blockchain networks, SBLWT’s private key isolation model, SPV verification within the Secure World, and secure UI design strategies can be leveraged. Moreover, SBLWT demonstrates the feasibility of modular TA design for handling the differing SPV structures and transaction signing requirements of various networks. 

Importantly, while Dai et al.’s work targeted a single blockchain network (Bitcoin), this research aims to extend the architecture to support dynamic firmware updates across multiple blockchain networks. Unlike SBLWT, this research envisions a server-driven process that dynamically composes optimized TA modules for different blockchain protocols, distributes them via OTA (Over-the-Air) updates, and securely applies them on OP-TEE-based platforms. 

Thus, Dai et al.’s SBLWT study represents a practical example of TEE-based lightweight blockchain wallet design and serves as an important technical foundation for this study’s multi-network, TEE-based wallet architecture.

\subsection{Security Aspects of Cryptocurrency Wallets — A Systematic Literature Review (Houy et al., 2023)}

\hspace*{1em}As blockchain technology has rapidly proliferated, cryptocurrency wallets have become essential tools for users to manage digital assets directly.\cite{paper_3} However, cryptocurrency wallets perform far more than simple key storage — they incorporate complex functionalities and are exposed to a wide attack surface with inherent structural vulnerabilities. Houy et al. (2023) addressed this reality by publishing a comprehensive systematic review on the security vulnerabilities and countermeasures of cryptocurrency wallets. 

The paper analyzed over 100 prior studies on cryptocurrency wallet security, providing a taxonomy of attack types and an overview of the entire threat landscape facing cryptocurrency wallets. Unlike earlier works that had been limited to analyses of key management, cryptographic modules, or specific protocols, this study provided a multilayered view of the complex functionality and attack surface unique to cryptocurrency wallets. 

Houy et al. defined cryptocurrency wallets as *“a composite security domain combining characteristics of password managers, banking applications, and privacy-focused systems,”* highlighting their broader security requirements beyond simple key storage. The paper analyzed wallet security from an attacker’s perspective across six major layers:

\begin{enumerate}[itemsep=0.5em, topsep=5pt, left=0pt]
    \item \textbf{Memory \& Storage} \\*Vulnerabilities include private key, seed phrase, and PIN theft through memory analysis and circumvention of encrypted storage.
    \item \textbf{Operating System} \\*OS-level attacks such as abuse of Android Accessibility features, USB debugging vulnerabilities, and installation of malicious apps.
    \item \textbf{Software Layer} \\*Application-level vulnerabilities including library weaknesses, bypassing RPC authentication, and leakage of sensitive information through logs.
    \item \textbf{Network Layer} \\*Threats such as Man-in-the-Middle (MITM) attacks, deanonymization via packet analysis, and BGP hijacking.
    \item \textbf{Blockchain Protocol} \\*Protocol-level attacks such as double-spending, 51\% attacks, malleability attacks, and eclipse attacks.
    \item \textbf{Other attacks} \\*Exchange hacks, side-channel attacks against air-gapped wallets, and other miscellaneous threats.
\end{enumerate}

The study empirically emphasized that OS and software-layer attacks represent the most critical real-world risks — underscoring the necessity of employing hardware-backed TEEs such as TrustZone to secure mobile cryptocurrency wallets. 

Houy et al. also reviewed the current state of TEE-based wallet research. Works such as Dai et al. (2018)’s SBLWT and Gentilal et al. (2017)’s TrustZone-backed Bitcoin Wallet were cited, but it was noted that these efforts largely focused on single blockchain networks (primarily Bitcoin) and lacked firmware-level update mechanisms or dynamic TA composition architectures. This finding strongly supports the differentiation and necessity of this study’s proposed architecture: a user-selectable blockchain wallet framework based on dynamic firmware updates and server-driven TA composition. 

Furthermore, the paper argued that as the industry moves toward a mainstream multi-chain environment, cryptocurrency wallets can no longer remain static — they must evolve to support dynamic architectures that adapt to different networks as selected by the user. 

This study’s proposed architecture — a server-based dynamic TA composition and firmware update process enabling OP-TEE-based wallets to adapt securely to user-selected blockchain networks — directly addresses the limitations highlighted by Houy et al., including:

\begin{itemize}[itemsep=0.0em, topsep=5pt, left=0pt]
    \item fixed single-network wallet designs,
    \item lack of dynamic updates, and
    \item insufficient flexibility for user-driven network selection.
\end{itemize}

Thus, Houy et al. (2023) serves as a key reference for justifying the multilayered security challenges faced by cryptocurrency wallets, and provides strong theoretical support for the need for firmware update and dynamic TA composition architectures — helping to clearly articulate the originality and necessity of this study compared to existing research.

\subsection{OP-TEE Official Documentation (v23.12.3)}

\hspace*{1em}Among open-source TEE platforms, OP-TEE is one of the most widely used, jointly developed by the open-source community, ARM, and Linaro. It is a standardized Trusted OS that enables Trusted Applications (TAs) to operate within the ARM TrustZone Secure World.\cite{paper_4} OP-TEE fully adheres to the GlobalPlatform (GP) TEE standards, and is regarded as one of the most practical and popular platforms for implementing and applying TrustZone-based TEEs in real-world applications. 

The OP-TEE Documentation (v23.12.3) serves as the official and comprehensive reference for OP-TEE architecture, functionalities, API usage, and system integration — a critical resource for TEE-based application development, system design, and kernel-to-user space integration.

This documentation provides an in-depth explanation of OP-TEE’s architecture, including core functionalities such as:

\begin{itemize}[itemsep=0.25em, topsep=5pt, left=0pt]
    \item Secure Monitor Call (SMC) flow
    \item Trusted Thread Scheduling
    \item Shared Memory management
    \item Secure Paging
    \item Crypto API architecture
    \item Secure Boot process
\end{itemize}

OP-TEE operates on a Dual World architecture, consisting of a Secure World and a Normal World. The Normal World typically runs a Linux-based OS (such as Android/Linux), while the Secure World runs the OP-TEE Trusted OS. This separation ensures that sensitive operations (such as key generation, signing, and decryption) are securely performed within the Secure World, protecting critical assets even against kernel-level compromises or malware in the Normal World. 

A key feature of OP-TEE is its management of SMC flow via the Secure Monitor Interface, with support for Yielding SMC-based Trusted Thread suspend/resume mechanisms. This feature is particularly important for this study’s proposed *server-based dynamic TA composition and firmware update* architecture, as it allows updated TAs to be applied without disrupting or conflicting with existing thread contexts. 

OP-TEE also supports non-contiguous Shared Memory management, enabling efficient communication with Linux kernel drivers and meeting diverse application requirements. In this research, where interaction with various blockchain-specific key-value storage structures is required, leveraging Shared Memory allows for optimized performance and data consistency. 

From a cryptographic perspective, OP-TEE supports a broad range of algorithms — including SHA, HMAC, ECDSA, RSA, and AES — via its GP TEE Cryptographic API, and also offers hardware crypto acceleration. As this research involves dynamic composition of TA modules to meet varying crypto requirements across blockchain networks (e.g., ECDSA P-256, Ed25519, KECCAK256), OP-TEE’s Crypto API abstraction layer provides a highly flexible and adaptable foundation.

Another noteworthy feature highlighted in the OP-TEE documentation is Secure Paging. Given the limited memory resources available in the Secure World, the Secure Pager allows efficient memory usage — a key capability for this research when flexibly loading and unloading multiple blockchain-network-specific TA modules. In the proposed architecture — *where TA modules are dynamically composed on the server, distributed as firmware updates, and securely applied via OP-TEE-based system updates* — the structural principles outlined in the OP-TEE documentation will be directly incorporated throughout the system design.

Specifically:

\begin{itemize}[itemsep=0.25em, topsep=5pt, left=0pt]
    \item Optimized communication between Linux user space and TAs via SMC-based RPC flow,
    \item Stable application of firmware updates through Trusted Thread Scheduling and suspend/resume mechanisms,
    \item Flexibility in composing network-specific TA modules through the Crypto API abstraction layer,
    \item High-performance data exchange via the Shared Memory architecture,
\end{itemize}

will all serve as essential technical components for building a flexible firmware update framework that supports multiple blockchain networks.

Ultimately, the OP-TEE documentation provides a systematic technical foundation for this study’s architecture. By adhering to GP standards and ensuring hardware portability, it will also enhance the portability, scalability, and general applicability of this research’s outcomes.

\subsection{TrustZone-backed Bitcoin Wallet (Gentilal et al., 2017)}

\hspace*{1em}In cryptocurrency wallets designed for securely managing blockchain-based digital assets, the protection of the private key remains the most critical security requirement.\cite{paper_5} However, software wallets operating in conventional mobile environments are vulnerable to OS-level compromises and are exposed to various attacks such as memory dumps, debugging, and hooking. While hardware wallets offer strong security guarantees, they also suffer from limited usability and require additional devices. 

To address these challenges, Gentilal et al. (2017) designed a TrustZone-backed Bitcoin Wallet (TBW) architecture that leverages ARM TrustZone-based Trusted Execution Environment (TEE) to enhance the core security functions of a Bitcoin wallet. 

This study presented a concrete implementation that mitigates the structural vulnerabilities of mobile wallets by moving private key protection, transaction signing, and cryptographic operations into the TrustZone-based Secure World.

The key features of the TBW architecture are as follows:

\begin{enumerate}[itemsep=0.5em, topsep=5pt, left=0pt]
    \item \textbf{Private key generation and storage performed within the Secure World} \\* The Normal World has no access to private keys, protecting them from memory analysis and OS-level attacks.
    \item \textbf{Transaction signing executed entirely in the Secure World} \\* Signature operations are performed exclusively in the Secure World, and only the signature output is exposed externally.
    \item \textbf{Random number generation conducted within the Secure World} \\* High-quality random number generation for ECDSA signing is securely implemented.
    \item \textbf{Persistent Secure Storage with Write Cache} \\* Performance degradation during key storage is minimized, ensuring efficient storage performance.
    \item \textbf{Standard cryptographic algorithms implemented using GP TEE Crypto APIs} \\* Algorithms such as ECDSA, AES, and SHA-256 are supported through standardized, secure interfaces.
\end{enumerate}

Gentilal et al. further highlighted that prior TrustZone-based solutions often relied on proprietary TEEs, limiting their applicability in open-source research. In contrast, TBW was implemented using OP-TEE, ensuring both platform independence and openness. Additionally, TBW was built upon the existing BitSafe platform, further reinforcing its security and portability. 

In terms of performance, the study demonstrated that despite performing cryptographic operations and storage tasks within the Secure World, key functions such as transaction signing and address generation exhibited improved performance. Latency increases from Persistent Secure Storage were effectively mitigated through the use of a Write Cache.

From a security perspective, TBW demonstrated superior results compared to conventional software wallets in the following areas:

\begin{itemize}[itemsep=0.25em, topsep=5pt, left=0pt]
    \item Enhanced resistance to dictionary attacks
    \item Improved side-channel attack resistance
    \item Secure Multi-TA architecture with robust isolation
\end{itemize}

The entire implementation was released as open source, making it available for other researchers to adopt.

In the context of this study’s proposed architecture — where TA modules are dynamically composed on the server, distributed as firmware updates, and securely applied through OP-TEE-based system updates — the TBW architecture serves as an important reference.

This study intends to actively apply TBW design principles such as:

\begin{itemize}[itemsep=0.25em, topsep=5pt, left=0pt]
    \item Migration of private key management and transaction signing to the Secure World
    \item Secure Storage combined with Write Cache strategies
    \item Standard Crypto API-based modular design
    \item Multi-TA architecture and inter-TA isolation policies
\end{itemize}

Furthermore, while TBW was designed for a single blockchain network (Bitcoin), this study targets dynamic support for multiple blockchain networks (Ethereum, Solana, EVM-compatible networks, etc.) by having server-side composed TA modules distributed via Over-The-Air (OTA) updates. The proposed architecture also aims to extend the security enhancements of TBW to various blockchain protocol characteristics. 

Additionally, it will incorporate Secure Paging and Yielding SMC-based thread management to ensure efficient operation of multi-network TA modules. 

Thus, Gentilal et al. (2017) represents a highly significant prior work, both in terms of practical implementation and secure design, providing valuable technical guidance for this study’s design of a flexible, multi-blockchain firmware update architecture.

\subsection{Conclusion}

\hspace*{1em}The prior works reviewed in this study have made substantial contributions to the design of TEE-based blockchain wallet architectures, particularly in areas such as private key protection, transaction signing integrity, implementation of lightweight SPV wallets, TrustZone-based security enhancements, and the establishment of standardized cryptographic operations.

However, several common limitations are evident across existing studies. First, most implementations are tied to specific blockchain networks, such as Bitcoin or Ethereum, and do not support dynamic selection by users or flexible support for multiple blockchain protocols.

Second, current TEE-based wallet architectures are generally static in design, with limited support for firmware-level updates or modular TA composition. They lack robust mechanisms to flexibly deliver new functionalities or security patches through OTA-based firmware updates.

Third, as multi-chain environments become increasingly prevalent in the Web3 era, there is a growing need for architectures that can quickly adapt to evolving network requirements and protocol changes through server-client coordinated TA composition and firmware updates. Existing studies have not adequately addressed this need for flexibility.

To overcome these limitations, this study proposes a novel architecture in which:

“When the user selects a blockchain network, the server dynamically composes the necessary TA modules and delivers them as a new firmware package, which is then applied via OP-TEE-based OS system updates.”

The goal of this architecture is to transcend the limitations of single-network static wallets and provide:

\begin{itemize}[itemsep=0.25em, topsep=5pt, left=0pt]
    \item User-driven flexibility in blockchain network selection,
    \item Server-side dynamic composition and distribution of TA modules,
    \item High compatibility and portability by leveraging OP-TEE-based TEE architecture and GP standards.
\end{itemize}

This approach aims to implement a next-generation TEE-based wallet architecture capable of securely and flexibly supporting multiple blockchain networks. Ultimately, it can evolve into a platform that meets both the security and user experience demands of the emerging Web3 and multi-chain era.

\section{Possible Threat Modeling}

\hspace*{1em}Blockchain wallets are critical components in the cryptocurrency ecosystem, responsible for securely managing users' private keys and enabling transactions. Despite their importance, these wallets face a wide range of security threats from both technical and human attack vectors. Understanding the diverse models of potential attackers is essential for developing more robust wallet designs and countermeasures. To this end, we provide a layered categorization of attacker models that target different aspects of wallet infrastructure, ranging from physical memory access to advanced protocol-level attacks. Table~5, presents a structured overview of these attacker models, grouped by layer—memory, operating system, software, network, protocol, and others—along with their respective objectives and notable references from the literature. This layered approach helps highlight how each component of the wallet stack can be targeted individually or in combination. The attacker models described go beyond theoretical concerns and are drawn from real-world incidents and research studies. Moreover, they demonstrate that even minor vulnerabilities at lower system levels can lead to catastrophic compromises of cryptographic assets. By identifying and understanding these threats in a structured manner, developers, auditors, and users can better anticipate risks. This analysis sets the foundation for evaluating wallet security holistically rather than focusing solely on the application logic or user interface.

\begin{table*}[ht]
\centering
\begin{tabular}{|
m{3.5cm}|
m{2cm}|
m{6.5cm}|
m{2.5cm}|
m{1.5cm}|}
\hline
\textbf{Type} & \textbf{Attacker Model} & \textbf{Description} & \textbf{Attacker's Goals} & \textbf{Reference} \\
\hline
\multirow{3}{*}{(a) Memory and Storage} 
& a-1 & 
Cold Boot Memory Extraction
& \multirow{3}{*}{Steal Creds} & \cite{halderman2008coldboot} \\
\cline{2-3} \cline{5-5}
& a-2 &
Memory/Storage Dump via Root
& & \cite{hu2021wallitiq} \\
\cline{2-3} \cline{5-5}
& a-3 & 
Wallet Encryption Brute-Force
&  & \cite{byun2024electronics} \\
\hline
\multirow{3}{*}{(b) Operating System} 
& b-1 & 
OS Kernel/Privilege Exploit
& \multirow{3}{*}{Steal Creds} & \cite{wallitiq2023whitepaper} \\
\cline{2-3} \cline{5-5}
& b-2 &
Abuse of OS Services (Accessibility)
& & \cite{leguesse2020wallets} \\
\cline{2-3} \cline{5-5}
& b-3 & 
Keylogging and Screen Capture
&  & \cite{wallitiq2023whitepaper} \\
\hline
\multirow{4}{*}{(c) Software Layer} 
& c-1 & 
Wallet Application Logic Flaw
& \multirow{4}{*}{Steal Creds} & \cite{erinle2025arxiv} \\
\cline{2-3} \cline{5-5}
& c-2 &
Privacy Leak in SPV Wallet
& & \cite{hu2021wallitiq} \\
\cline{2-3} \cline{5-5}
& c-3 & 
Unsolicited Transaction Spam
&  & \cite{hu2021wallitiq} \\
\cline{2-3} \cline{5-5}
& c-4 & 
Trojan or Malicious Wallet App
&  & \cite{techxplore2021wallets} \\
\hline
\multirow{4}{*}{(d) Network Layer} 
& d-1 & 
P2P Deanonymization Attack
& \multirow{4}{*}{\shortstack{Deanonymize\\ / Steal Creds}} & \cite{biryukov2014deanonymization} \\
\cline{2-3} \cline{5-5}
& d-2 &
Eclipse Attack
& & \cite{heilman2015eclipse} \\
\cline{2-3} \cline{5-5}
& d-3 & 
BGP/Routing Hijack
&  & \cite{apostolaki2017hijacking} \\
\cline{2-3} \cline{5-5}
& d-4 & 
DNS Hijacking / MITM on API
&  & \cite{wallitiq2023whitepaper} \\
\hline
\multirow{3}{*}{(e) Blockchain Protocol} 
& e-1 & 
Transaction Double-Spending (Fast Payments)
& \multirow{3}{*}{Create Coins} & \cite{karame2012doublespending} \\
\cline{2-3} \cline{5-5}
& e-2 &
Consensus Majority Attack (51\% Attack)
& & \cite{mnakamoto2008bitcoin} \\
\cline{2-3} \cline{5-5}
& e-3 & 
Cryptographic Breaks (Weak RNG/Key Reuse)
&  & \cite{buterin2013ethereum} \\
\hline
\multirow{4}{*}{(f) Other} 
& f-1 & 
Social Engineering \& Phishing 
& \multirow{4}{*}{\shortstack{Deanonymize\\ / Steal Creds}} & \cite{vasek2015scams} \\
\cline{2-3} \cline{5-5}
& f-2 &
Insider Threat (Custodial Wallets)
& & \cite{erinle2025arxiv} \\
\cline{2-3} \cline{5-5}
& f-3 & 
Physical Coercion (“\$5 Wrench Attack”)
&  & \cite{trdcrft2024wrench} \\
\cline{2-3} \cline{5-5}
& f-4 & 
Hardware Side-Channel Attack
&  & \cite{genkin2015ecdsa} \cite{halborn2021hardwarewallet} \\
\hline
\end{tabular}
\vspace{1mm}
\caption*{\textbf{Table 5: Overview of Attacker Models}}

\vspace{-3mm}
\caption*{\textit{Table: Overview of threat models for blockchain wallets, categorized by layer. Each entry includes the attack vector, attacker’s objectives, and references to research.}}
\end{table*}

\subsection{(a) Memory and Storage Threats}
\hspace*{1em}Memory and storage attacks focus on extracting or compromising secret keys from the wallet’s device memory or persistent storage. Cold boot attacks are a prime example: an attacker with physical access can freeze the device’s RAM chips to slow memory decay, then reboot or transplant them to recover whatever was in memory (including private keys)(a-1). Similarly, if a device is rooted or infected with malware, an attacker can dump memory or read files to obtain unencrypted wallet(a-2). Even encrypted wallets are vulnerable if protected by weak passwords – attackers can perform brute-force attacks on wallet files, exploiting the relatively short or low-entropy passwords allowed by some wallet(a-3). In all these cases, the attacker’s goal is to steal the private keys or seed phrases that unlock the victim’s funds, thereby gaining control over the cryptocurrency.

\subsection{(b) Operating System Threats}
\hspace*{1em}Operating system threats leverage weaknesses in the platform running the wallet. If the OS itself is compromised, the wallet’s security can be bypassed. For instance, an attacker might exploit an OS vulnerability to gain root privileges, effectively removing all isolation – once they have root, they can directly read or modify wallet processes and files(b-1). Attackers also abuse legitimate OS features: on Android, malware can misuse Accessibility services to invisibly observe and interact with apps. Research has demonstrated a stealthy accessibility trojan that can take over a popular mobile wallet app’s UI, auto-confirming transactions (even intercepting 2FA prompts) to withdraw funds without the user’s consent(b-2). Furthermore, keylogging or screen-capture malware operating at the OS level can record everything the user types or sees on their screen(b-3) – for example, capturing a seed phrase as it’s being displayed, or the PIN as it’s entered. The goals of OS-layer attacks are to get around the wallet’s application-level security by exploiting the underlying system, allowing attackers to steal keys or spoof user actions (like transaction approvals).

\subsection{(c) Software Layer Threats}
\hspace*{1em}Software layer threats target vulnerabilities in the wallet application itself (or its supporting libraries). These can range from classic software bugs to design flaws specific to crypto wallets. For instance, a logic error in how a wallet verifies transaction signatures or permissions could let an attacker bypass authentication and spend funds they don’t own(c-1). One well-documented case is the BitcoinJ library vulnerabilities affecting many Bitcoin wallet apps: a flaw caused SPV (simplified payment verification) wallets to leak all their Bitcoin addresses, enabling a network eavesdropper to deanonymize users(c-2). Another bug in the same library made wallets download excessive transaction data in the background, which attackers turned into a spamming attack to drain victims’ phone batteries and data plans(c-3). Beyond unintended bugs, there’s also the risk of malicious wallet software: for example, attackers have created fake wallet apps that look legitimate but contain backdoors. Unsuspecting users who install these essentially hand their private keys to the attacker. As one researcher noted, users should be cautious because any developer (even a malicious one) can publish a wallet app on official app stores(c-4). In summary, software-layer attacks aim to exploit or insert vulnerabilities in the wallet program, either to undermine its security controls (stealing keys, bypassing checks) or degrade its operation (denial of service).

\subsection{(d) Network Layer Threats}
\hspace*{1em}
Network threats involve the wallet’s communication with the outside world – typically the blockchain network or backend servers. One major category is privacy attacks: by observing the wallet’s network traffic, attackers can perform network-layer deanonymization. For example, analysis of Bitcoin’s peer-to-peer network has shown that it’s possible to link a user’s Bitcoin addresses to their IP address by mapping how transactions propagate(d-1). This breaks the pseudonymity of the wallet, potentially exposing the user’s identity and transaction history. Other network attacks are more active. In an eclipse attack, an adversary controls enough nodes to isolate a wallet’s network connections entirely(d-2). The wallet becomes trapped behind a set of attacker-controlled peers, causing it to accept a false version of the blockchain or miss critical updates. An eclipsed wallet can be fed double-spend transactions or prevented from seeing that it’s underpaid, facilitating fraud. On a broader scale, attackers can exploit Internet routing – BGP hijacking can divert the wallet’s traffic to malicious routes(d-3). This might partition the wallet from the true network or allow the attacker to interpose themselves as a “man in the middle.” For instance, research shows that BGP-based partitioning can significantly delay block propagation and even enable confirmed double-spending. Finally, there are DNS hijacking and MITM attacks targeting web-based wallets or APIs. If a wallet relies on a domain (for fetching balance info or sending transactions), an attacker who hijacks that domain’s DNS can redirect the wallet to a fake server(d-4). In practice, this has led to phishing pages that trick users into entering their private keys. Likewise, an attacker on the same network (e.g., a public Wi-Fi) could intercept API calls if they’re not encrypted, altering destination addresses in transit. Network-layer attackers typically aim to either deanonymize the user or manipulate the data in transit, with outcomes like privacy loss, fraudulent transactions, or denial-of-service.

\subsection{(e) Blockchain Protocol Threats}
\hspace*{1em}
Protocol-level threats stem from the underlying rules and design of the blockchain system that the wallet operates on. One classic issue is lack of immediate transaction finality, which enables double-spending attacks on fast payments. If a merchant’s wallet accepts a payment on zero or few confirmations (for speed), an attacker can exploit that by quickly broadcasting a conflicting transaction (or even controlling mining of a couple of blocks) to invalidate the first transaction(e-1). The result is the merchant sees a payment that later disappears from the ledger, letting the attacker get goods for free. At a deeper level, if attackers gain control of the consensus process – notably, more than 50\% of the network’s mining or staking power – they can mount a 51\% attack. In this scenario, the adversary can rewrite the blockchain history at will(e-2). That means they could reverse even long-confirmed transactions (perhaps double-spending high-value transfers back to themselves) and censor new transactions from being included. Such majority attacks have moved from theory to reality on smaller blockchains, leading to significant losses. Another aspect of protocol security is the strength of the cryptographic algorithms. Wallet security can be totally undermined by cryptographic vulnerabilities – for example, the infamous 2013 incident where Android’s flawed random number generator caused wallets to reuse ECDSA signature nonces, allowing attackers to compute the private keys(e-3). Similarly, if an attacker ever found a break in the hash function or elliptic curve used (or had a quantum computer), they could forge transactions or steal funds directly. In summary, protocol-layer threats involve attacking the rules or math of the system itself – either by exploiting the way confirmations work (as in double spends and consensus attacks) or by subverting the cryptographic assumptions (as in weak RNGs or future quantum attacks).

\subsection{(f) Other Threats}
\hspace*{1em}
This category includes threats that don’t neatly fit into the technical layers but are nonetheless significant. Social engineering is a major one: instead of hacking software, attackers hack the user’s trust and habits. Phishing emails, fake support calls, or fraudulent websites can convince users to divulge their seed phrase or enter credentials on a malicious site(f-1). Once the attacker has this information, they don’t need to break any encryption – they simply log in or import the wallet and drain the funds. Another “out-of-band” threat is the insider threat when using custodial wallets or exchanges. In these scenarios, the user’s keys are held by a service. A dishonest employee or compromised insider can directly access many users’ keys or the pooled funds. There have been cases where large exchanges were looted, possibly with insider assistance(f-2). The attacker’s motive here is straightforward financial gain, but the attack bypasses all technical defenses by exploiting internal access. Additionally, physical coercion attacks are a real-world threat: a robber might target a crypto holder, forcing them at gunpoint to unlock a hardware wallet or reveal a PIN(f-3). This is often grimly referred to as the “five-dollar wrench attack,” implying that an inexpensive wrench (used as a weapon) can negate even the most advanced digital security. Finally, on the more exotic end, side-channel attacks on hardware wallets are a concern(f-4). These devices are engineered to be secure, but researchers have shown that monitoring things like power consumption or electromagnetic emissions during device operation can leak secrets. For instance, an attacker with the right equipment could potentially extract a PIN or even a private key from a hardware wallet without physically opening it, by analyzing its power usage patterns during a signing operation. While these attacks require sophistication and sometimes close proximity, they highlight that no system is completely immune. In essence, the “Other” category reminds us that human factors, physical security, and hardware quirks are all part of the threat landscape for blockchain wallets, beyond just software and network vulnerabilities.

\vspace{0.5\baselineskip}

\indent The threat landscape facing blockchain wallets is both multifaceted and evolving, encompassing low-level hardware attacks, OS-level exploits, software vulnerabilities, and sophisticated social engineering techniques. As shown in Table~5, attackers can leverage a wide spectrum of methods to extract private keys, deanonymize users, or bypass wallet protections entirely. Importantly, no single security mechanism is sufficient on its own; layered defense strategies must be adopted to account for diverse attacker capabilities. Memory and storage vulnerabilities highlight the importance of physical security and encryption, while OS- and software-layer threats stress the need for sandboxing and code audits. Network-layer exploits and blockchain protocol attacks underscore the significance of robust cryptographic design and secure communication. Beyond technical attacks, human-centered threats such as phishing and coercion remind us that end-users are often the weakest link in the security chain. These findings reinforce the necessity of considering attacker goals, system layer interactions, and real-world constraints when designing secure wallets. Future wallet implementations must prioritize modularity, transparency, and resilience to emerging threats. Ultimately, recognizing these attack models equips stakeholders with the foresight to mitigate vulnerabilities before they are exploited. Through this detailed mapping of attacker models, we aim to contribute to more secure and trustworthy blockchain wallet ecosystems.

\section{Design}

\hspace*{1em}As the adoption of blockchain technologies continues to grow, the importance of securing digital assets stored in cryptocurrency wallets has become increasingly critical. Conventional software wallets—while convenient—expose users to a variety of threats ranging from malware in the operating system to side-channel attacks that can extract sensitive information such as private keys. To mitigate such risks, hardware-backed solutions such as Trusted Execution Environments (TEEs) have emerged as a promising approach to isolating sensitive operations and data from potentially compromised system components.

This paper presents the design of a modular, TEE-based blockchain wallet platform that leverages ARM TrustZone technology to protect cryptographic operations and key materials within a secure execution context. By splitting the system into two distinct domains—the Rich Execution Environment (REE) and the Trusted Execution Environment (TEE)—the platform ensures that sensitive tasks like key generation and transaction signing are handled by isolated Trusted Applications (TAs) running in the secure world, while the user interface and network interactions remain in the REE.

A key advantage of this design lies in its modularity: each supported blockchain network is encapsulated within its own set of TAs, enabling easy extensibility and minimizing the trusted computing base for any single operation. The system also introduces a secure firmware update mechanism that allows dynamic deployment of blockchain-specific TAs, thereby maintaining a minimal attack surface and enabling scalable support for diverse cryptocurrency protocols.

Through a detailed architectural breakdown, secure workflow analysis, and discussion of TA composition, firmware management, and security considerations, this paper demonstrates how a well-structured TEE-based platform can offer both strong protection and practical adaptability for secure blockchain wallet implementations on modern Android devices.

\begin{figure*}[ht]
    \centering
    \includegraphics[width=0.85\linewidth]{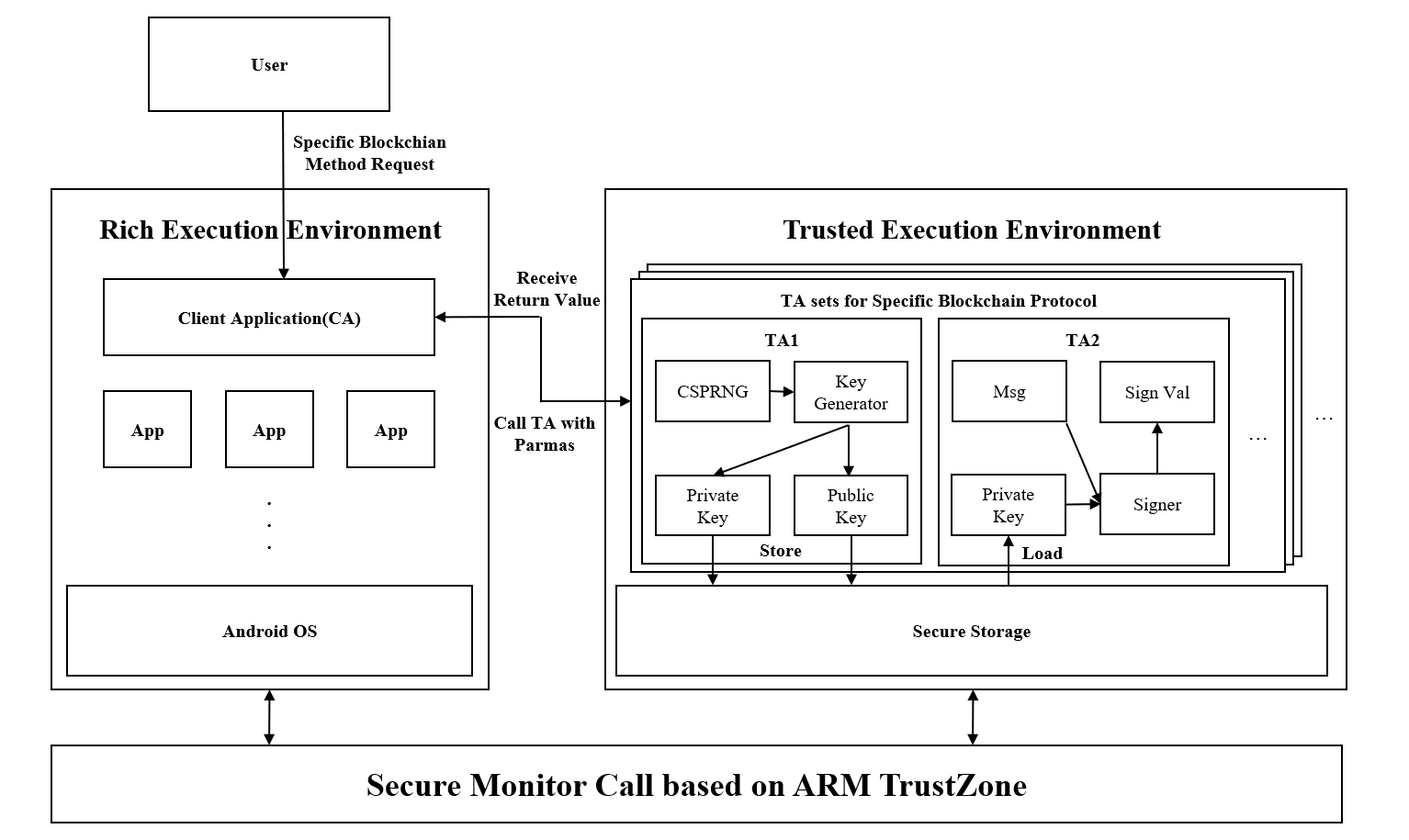}
    \caption{Overall architecture of the TEE-based wallet}
    \label{fig:architecture}
\end{figure*}

\subsection{Architecture Overview}

\hspace*{1em}Figure ~5 shows the architecture of the proposed TEE-based blockchain wallet platform. The system is divided into two distinct domains: the Rich Execution Environment (REE) and the Trusted Execution Environment (TEE). The REE, represented by the Android OS and normal applications, is where the user interacts with the system through a Client Application (CA). This client app runs in the normal (untrusted) world and acts as a bridge between the user and the secure services. The TEE, leveraging ARM TrustZone technology, runs in an isolated secure world (e.g., using OP-TEE as the TEE operating system) and hosts specialized Trusted Applications (TAs) that handle sensitive operations like cryptographic key management and transaction signing. All normal applications, including the wallet’s CA and other apps, operate in the REE, which has rich functionality but is considered untrusted from a security perspective. In contrast, the TEE is a tightly controlled environment that ensures confidentiality and integrity for code and data within it.

In this architecture, the CA in the REE issues requests to the TEE whenever a security-critical operation is needed. The boundary between REE and TEE is enforced by hardware; any interaction crossing this boundary (from CA to TA and back) occurs through a Secure Monitor Call (SMC) facilitated by the TrustZone secure monitor. The CA does not directly access secure resources; instead, it packages the user’s request and invokes the corresponding TA in the TEE via the secure TEE driver interface. The TEE’s operating system receives the request, identifies the target TA (one of the blockchain-specific modules), and passes along the request parameters. The figure shows multiple TAs grouped as “TA sets” for a specific blockchain protocol—highlighting that for each supported blockchain network, the platform includes a dedicated set of TAs to perform that chain’s cryptographic functions. A secure storage component resides in the TEE to hold private keys and other sensitive data. The Android OS in the REE cannot directly read or modify this secure storage due to the privilege separation enforced by TrustZone. Overall, this architecture ensures that all critical cryptographic operations and key material remain within the protected TEE, while the REE handles user interface and network communication aspects of the wallet.

\subsection{TEE Workflow}

\hspace*{1em}When the user initiates a sensitive operation (such as generating a new key pair or signing a transaction) through the wallet’s client application in the REE, the request is securely handled through the following steps:

\begin{enumerate}[itemsep=0.5em, topsep=5pt, left=0pt]
    \item \textbf{User Request:} The user triggers a blockchain-related request via the Client Application (e.g., tapping a “Create Key” or “Sign Transaction” button in the app).
    \item \textbf{CA Invokes TA:} The Client Application packages this request (including any necessary data, like transaction details) and invokes the appropriate TA in the TEE. This invocation uses the device’s TEE driver/API, which under the hood triggers an SMC instruction to switch the context from the normal world (REE) to the secure world (TEE).
    \item \textbf{Dispatch to TA:} Upon entering the TEE, the secure world OS (OP-TEE) dispatches the request to the targeted Trusted Application. For instance, if the user requested a new key, the call is routed to the Key Generation TA (TA1); if the user requested a signature, it is routed to the Signing TA (TA2).
    \item \textbf{Secure Processing:} Inside the TA, the requested operation is executed in isolation. For key generation, TA1 utilizes a cryptographically secure random number generator (CSPRNG) to generate a new private–public key pair. The private key is stored in the TEE’s secure storage, and the public key (or a derived address) is returned. For a signature request, TA2 retrieves the corresponding private key from secure storage, uses it to compute the digital signature on the provided message or transaction hash, and then outputs the signature.
    \item \textbf{Return to REE:} The TA completes the operation and returns the result (e.g., the newly generated public key or the transaction signature) back to the REE via the secure monitor. The result is passed to the waiting Client Application.
    \item \textbf{Response to User:} The Client Application receives the result and can then proceed to use it in the regular workflow. For example, it may display a new public address to the user, or attach the signature to a transaction and broadcast it to the blockchain network. Throughout this process, the private key and any sensitive computations never leave the TEE; they remain protected by hardware isolation and are invisible to the REE.
\end{enumerate}

This workflow ensures that even if the REE (Android OS or the client app) were to be compromised by malware, the attacker cannot extract keys or forge signatures, because all critical operations are executed in the secure world by trusted code.

\subsection{TA Composition}

\hspace*{1em}To accommodate different blockchain protocols and maintain a clean separation of concerns, the platform’s secure logic is divided into multiple Trusted Applications. Each supported blockchain network is implemented with a specific set of TAs, collectively referred to as a “TA set” for that blockchain. These TAs are modular components within the TEE, each responsible for a distinct aspect of the wallet’s functionality. By dividing the functionality, the design promotes modularity, ease of maintenance, and security through least privilege (each TA only does what it needs to and has limited scope).

The core TAs in a typical blockchain TA set include:

\begin{itemize}[itemsep=0.5em, topsep=5pt, left=0pt]
    \item \textbf{TA1 – Key Generation:} This TA is responsible for creating new cryptographic key pairs for the blockchain. It uses a CSPRNG in the secure world to generate a private key, derives the corresponding public key, and securely stores the private key (for example, in the TEE’s secure storage, possibly indexed by a key ID or associated with the blockchain account). It may return the public key or an address derived from it to the REE so the user can use it as their wallet address on the blockchain. All cryptographic computations for key generation occur inside TA1, ensuring the private key is never exposed to the REE.

    \item \textbf{TA2 – Transaction Signing:} This TA handles digital signing operations for transactions or messages. When the user needs to authorize a blockchain transaction, the client app invokes TA2 with the transaction data (or a hash of the transaction). TA2 then loads the appropriate private key from secure storage (the key that was previously generated by TA1 for this blockchain account), and uses it to produce a signature using the blockchain’s required cryptographic algorithm (e.g., ECDSA or EDDSA, depending on the blockchain). The resulting signature is returned to the REE so that the client app can attach it to the transaction and send it to the blockchain network. At no point does the private key leave TA2 or the secure world—only the signature is output.

    \item \textbf{Additional TAs:} If a particular blockchain protocol requires other specialized secure operations, additional TAs can be added to that blockchain’s TA set. For example, a blockchain might need a TA for a specialized cryptographic operation or protocol-specific key derivation beyond simple signing. Each TA is kept independent and focused on a specific task. This modular approach means new TAs can be introduced or updated without affecting the rest of the system, as long as they adhere to the defined interfaces with clean encapsulation.
\end{itemize}

Each TA is isolated from others by the TEE’s runtime, which means a vulnerability in one TA (if ever exploited) should not directly compromise the code or data of other TAs. The use of distinct TAs per functionality and per blockchain also allows the platform to include support for a new blockchain by adding a new set of TAs, rather than altering existing trusted code. This structure is key to the platform’s flexibility in supporting multiple cryptocurrencies securely.

\begin{figure*}[ht]
    \centering
    \includegraphics[width=0.85\linewidth]{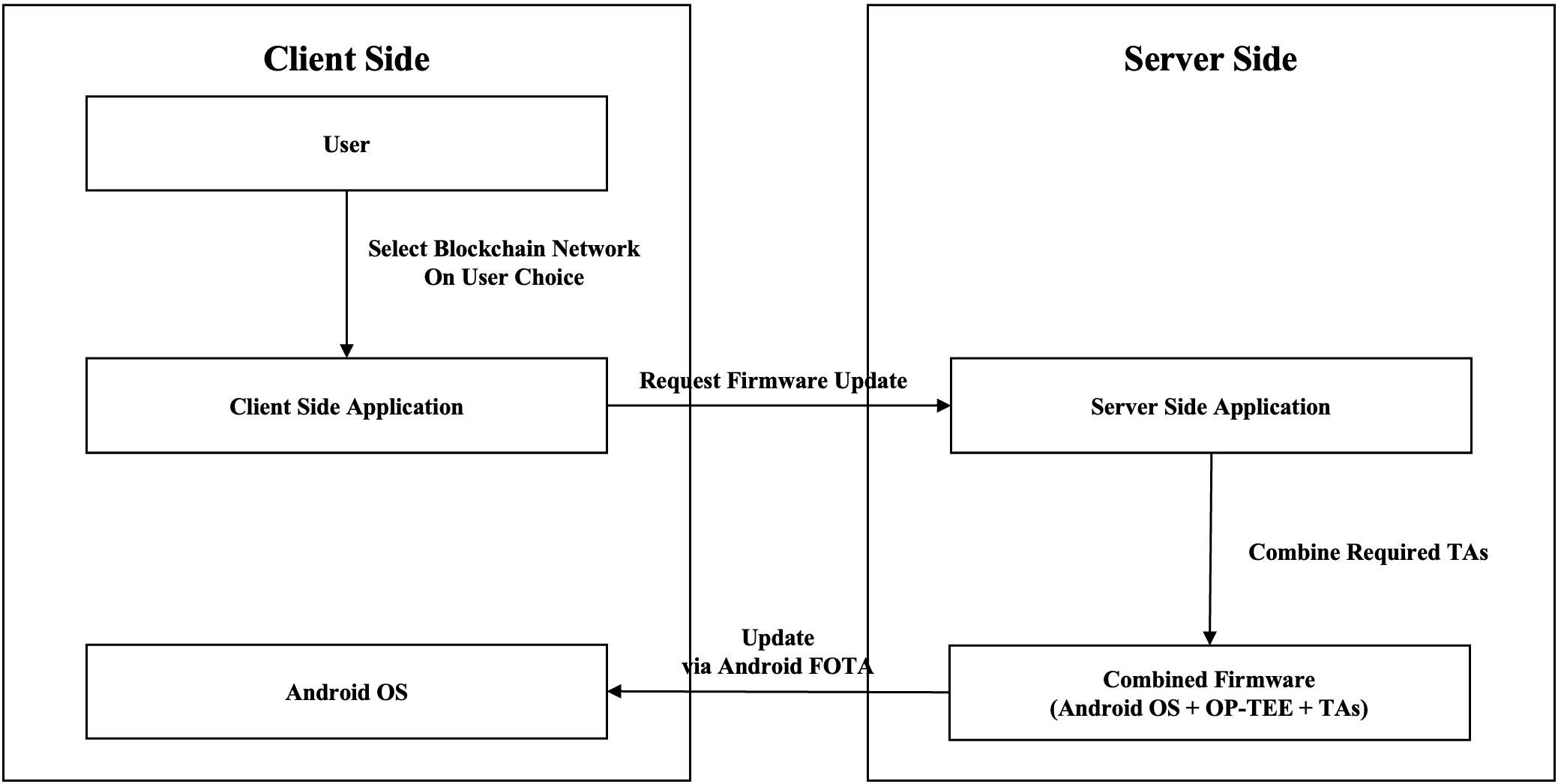}
    \caption{Overall architecture of the platform update method}
    \label{fig:architecture}
\end{figure*}

\subsection{Firmware Update Mechanism}

\hspace*{1em}Figure ~6 shows the firmware update mechanism for deploying blockchain-specific TAs on the device. Supporting multiple blockchain networks in a secure manner is achieved by dynamically updating the device’s firmware to include the necessary TEE components for the chosen network. Rather than pre-loading all possible blockchain TAs (which would increase the attack surface and device footprint), the platform delivers targeted updates that package only the required secure components.

The update process works as follows:

\begin{enumerate}[itemsep=0.5em, topsep=5pt, left=0pt]
    \item \textbf{Network Selection:} On the client side, the user selects the desired blockchain network through the wallet’s interface (for instance, choosing between Bitcoin, Ethereum, etc.). This choice indicates which blockchain’s support the user wants to activate on the device.
    \item \textbf{Update Request:} The Client Application (running in Android) sends a firmware update request to the platform’s server infrastructure, specifying the chosen blockchain network. This request can be made over a secure internet connection to a designated Server Side Application.
    \item \textbf{Preparing Combined Firmware:} On the server side, the system prepares a combined firmware image tailored for the requested blockchain. This combined firmware includes three main parts: the Android OS update (if needed), the TEE component (OP-TEE) configured for that device, and the set of TAs required for the selected blockchain. Essentially, the server “combines” the necessary TAs with the base firmware. If the device already has the latest Android OS, the focus might just be on updating the TEE and adding new TAs. All code (especially the TAs and TEE changes) is signed or verified by the platform to ensure its integrity and authenticity.
    \item \textbf{FOTA Delivery:} The server then delivers this combined firmware package to the client device using Android’s Firmware Over-The-Air (FOTA) update mechanism. The device receives the update package and, upon user consent, proceeds to install it. The update process leverages Android’s standard update flow, which typically involves verifying the package’s cryptographic signature, then rebooting the device into a recovery or update mode to apply the new firmware.
    \item \textbf{Integration of TAs:} Once the update is applied and the device reboots, the new firmware becomes active. The Android OS is now coupled with an updated OP-TEE environment that contains the newly added TA set for the requested blockchain. For example, if the user selected a certain blockchain, the corresponding TA1, TA2 (and any additional TAs for that blockchain) are now installed in the TEE.
    \item \textbf{Wallet Ready:} After the update, the wallet application on the device can detect that the new TAs are available (sometimes the app might verify the presence or version of the TA set). The user can now use the wallet app to perform operations on the newly supported blockchain, with all security guarantees in place. Should the user want to support a different blockchain later, the process can be repeated to update the firmware with a different set of TAs.
\end{enumerate}

This firmware update mechanism ensures that the device always runs a minimal set of trusted code in the TEE – only what is necessary for the blockchains the user actively uses. It balances security (smaller attack surface and up-to-date trusted code) with flexibility (the ability to support many blockchains on the same device via on-demand updates). Utilizing Android’s built-in FOTA system provides a trusted and user-friendly way to deliver these security-critical updates.

\subsection{Security Considerations}

\hspace*{1em}The design leverages ARM TrustZone-based isolation to achieve a strong security posture for the wallet. By segregating the execution environments, the platform ensures that even a compromised Android OS or malicious app in the REE cannot directly harm the TEE or extract sensitive data from it. The hardware-enforced memory separation means that secrets like private keys, which reside in secure world memory or storage, are inaccessible from the normal world. Only the secure monitor (via SMC calls) can mediate interactions, and the TEE will only perform operations for authorized requests coming from the legitimate CA.

\begin{itemize}[itemsep=0.5em, topsep=5pt, left=0pt]
    \item \textbf{Secure Storage:} One of the key security features is the use of secure storage in the TEE for private keys and other sensitive information. When a key pair is generated by TA1, the private key is stored in encrypted form within the TEE’s storage (often tied to hardware unique keys or the TEE’s root of trust). This means that even if an attacker obtains physical access to the device’s storage or if the Android OS is rooted, the raw private key material cannot be retrieved or decrypted without going through the TEE. The secure storage, combined with TEE’s enforcement that only the TA that created or is authorized to access that data can retrieve it, adds a robust layer of protection for the user's cryptographic keys. In our design, TA2 can load the private key only via the secure TEE APIs, ensuring that the key never leaves the secure world in plaintext form.
    \item \textbf{Isolation of TAs:} Each TA runs in the context of the TEE where the TEE OS (OP-TEE) provides isolation between different TAs. This means that even within the secure world, TAs are compartmentalized to some extent. A vulnerability in one TA (for example, a bug in a blockchain-specific TA) should not easily allow an attacker to breach another TA or the TEE kernel itself. The principle of least privilege is applied: each TA is given only the permissions and data access it requires to perform its function. For instance, TA2 might only have access to the specific key objects it needs for signing, and not to other keys belonging to different blockchains or accounts. This minimizes the damage that any single compromised TA could inflict.
    \item \textbf{Trusted Firmware Updates:} The firmware update mechanism itself is designed with security in mind. By using Android’s FOTA, the platform relies on established secure update practices: the update packages are digitally signed by the device/vendor’s private key, and the device will verify this signature before installing. This prevents malicious or tampered firmware from being installed. Thus, when adding new TAs via an update, users (and the device) can trust that the source is authentic and the code has not been altered in transit. Additionally, because the updates include the entire TEE stack, there is less risk of partial updates or mismatched versions; the TEE and TA code are tested as an integrated firmware image.
    \item \textbf{TEE Root of Trust:} The security model assumes a hardware root of trust. TrustZone ensures that on boot, the secure world (including OP-TEE and the pre-installed TAs) is initialized before the normal world, and that the bootloader verifies signatures on the secure world components. This secure boot process means the TEE code hasn’t been tampered with from power-on. Combined with the runtime isolation and secure update process, the wallet platform maintains a chain of trust from boot to runtime operations.
\end{itemize}

Overall, the platform’s security is anchored in keeping all private keys and cryptographic operations inside the TEE. Even if an attacker controls the REE, they cannot directly extract keys or make the TEE sign arbitrary data without going through the proper interfaces (which include necessary authorization, such as the user’s action in the app). This significantly reduces the attack surface compared to a conventional software wallet that stores keys in the normal environment. The use of multiple TAs further compartmentalizes the trust boundaries, and the secure update mechanism ensures the system can be kept up-to-date against vulnerabilities or expanded to support new algorithms without compromising security.

\subsection{Modularity and Extensibility}

\hspace*{1em}A central benefit of this TEE-based design is its modularity and extensibility to support a wide range of blockchain networks. Because the functionality for each blockchain is encapsulated in its own TA set, new blockchains can be added by developing and deploying new TAs rather than redesigning the entire system. This modular approach makes the platform future-proof and scalable:

\begin{itemize}[itemsep=0.5em, topsep=5pt, left=0pt]
    \item \textbf{Independent Blockchain Modules:} Each blockchain’s logic (key handling, signing, etc.) lives in its dedicated TAs. This independence means that adding support for a new blockchain is as simple as writing the appropriate TAs for that chain’s cryptographic algorithms and transaction formats, and then delivering them via a firmware update. Existing TAs for other blockchains remain unchanged and unaffected. The client application can be designed to recognize which blockchains are available (perhaps through a registry of installed TAs or a version query to the TEE) and present the appropriate options to the user.
    \item \textbf{Major vs. Minor Blockchain Support:} The platform architects can focus on building TAs in-house for major, widely-used blockchains to ensure optimal security and performance for those. For more niche or emerging blockchains (“minor” chains), the platform can allow third-party vendors or the blockchain’s developers to supply their own TAs that plug into the system. Such vendor-supplied TAs would still need to be vetted and signed by the platform maintainers to be accepted in the firmware update process, but this approach distributes the development effort. It allows the wallet to support many different cryptocurrencies without the core team having to implement every single one, thus encouraging a broader ecosystem of support.
    \item \textbf{Extensibility Through Updates:} Whenever a new TA or an updated version of an existing TA is available (be it to support a new blockchain or to enhance/patch an existing one), the firmware update mechanism can be utilized to deploy it to users’ devices. This ensures that the platform can evolve over time – new cryptographic algorithms or blockchain forks can be accommodated by updating the TAs. Because each TA is a self-contained module, updates or additions usually do not require changes to the rest of the system. For instance, if a new signature scheme needs to be supported for a particular blockchain, a new version of that blockchain’s signing TA can be deployed without touching the key generation TA or any other blockchain’s TAs.
    \item \textbf{Android-Specific Design:} This entire design is built for Android devices, leveraging Android’s update system and the widespread availability of TrustZone in modern ARM-based smartphones. By focusing on Android, the platform can integrate tightly with the Android OS (for example, using Android system services to trigger FOTA updates or using Android permissions and keystores as an additional layer of security for the CA). While the concept could in theory be extended to other operating systems that support a TEE, Android provides a standardized environment where this design can be implemented and deployed on a variety of hardware. This makes the solution practical for real-world use, as many manufacturers and users are familiar with FOTA updates and Android’s security model.
\end{itemize}

In summary, the design emphasizes security and modularity. It ensures that the wallet is secure by design (thanks to TrustZone isolation and careful separation of duties among TAs) and is also adaptable to the fast-evolving blockchain landscape. New features or blockchain supports can be integrated with minimal friction, simply by adding or updating isolated TAs, all while preserving the overall security of the system.

\section{Future Works}

\subsection{Prototype Implementation}

\subsubsection{Test on the Open-TEE}

Open-TEE is an open-source implementation of a “virtual TEE” (Trusted Execution Environment) that fully complies with the GlobalPlatform TEE specifications\cite{openTEE2025docs}. According to the official Open-TEE documentation, it is intended primarily as a development and testing tool for Trusted Applications: it enables developers and researchers to build and debug TEE-based systems even without access to hardware-based TEEs\cite{openTEE2025docs}. The Open-TEE project further notes that Trusted Applications developed with this virtual TEE can be compiled and run on any target that adheres to the same specifications\cite{openTEE2025docs}. In our work, we followed the official Open-TEE build instructions and successfully compiled the Open-TEE engine (open\_engine) and sample Trusted Applications in a QEMU-based ARM environment(Ubuntu 20.04 on VMware Workstation 17). After launching the engine under QEMU, we verified that the tee\_manager and tee\_launcher processes started correctly and that the provided conn\_test\_app client executed without errors\cite{openTEE2025docs}.

As a concrete demonstration, we implemented a Trusted Application within Open-TEE that performs Bitcoin transaction signing using ECDSA on the secp256r1 (NIST P-256) curve. This TA uses the Open-TEE Crypto API: it initializes an elliptic-curve key pair, computes the SHA-256 digest of the input transaction, and then generates an ECDSA signature with the private key. The normal-world client application invokes this TA via the Open-TEE interface, and we confirmed that the TA produced valid signatures for our test vectors. All source code developed in this project is publicly available on GitHub\cite{mm0ck3r2025opentee}.

\subsubsection{Android Build and Vendor Customization}

We also built the Android OS with OP-TEE integration in a QEMU environment. Using the official AOSP build process, we initialized the Android source repository and configured the build environment (for example, by running source build/envsetup.sh). We then selected a QEMU-based target that includes TEE support. For reference, the official Android documentation shows how to build Android with Trusty support by using a lunch target such as qemu\_trusty\_arm64-userdebug and then running make\cite{android2025downloadandbuild}. By analogy, we created a corresponding lunch target for OP-TEE and invoked the build, producing a complete Android system image with the OP-TEE kernel drivers and libraries included.

On the vendor side, we integrated the OP-TEE Linux kernel driver and the OP-TEE client libraries into the Android build. This involved adding the necessary OP-TEE kernel modules (for example, optee.ko) and the OP-TEE client binaries to the vendor partition, and modifying the device tree and kernel configuration to enable OP-TEE support. These vendor-specific modifications follow the general guidelines in Android’s official documentation for adding new hardware and services. In our experiments, the modified Android image booted correctly in QEMU with the OP-TEE driver loaded, and we were able to run simple trusted service tests (for example, invoking a test TA) on the emulated device.

These platform changes are directly relevant to future secure wallet applications. The Android documentation emphasizes that the presence of a TEE in the system provides hardware-backed strong security services to Android and its applications\cite{android2025keystore}. In particular, Android’s keystore architecture relies on a TEE-based Trusted Application (KeyMint TA) to perform all sensitive cryptographic operations\cite{android2025keystore}. By enabling OP-TEE support at the platform level, our modifications ensure that cryptographic services such as secure key storage and digital signing can be executed in a trusted environment. This paves the way for wallet applications to use the Android TEE for secure key management and transaction signing in future work.

\subsection{Overview}
\hspace*{1em}
This study demonstrated the design validity of the modular wallet and the functional and security potential of an initial prototype, but there remain numerous tasks to be solved for a production-level deployment. This section concretizes the future research and development roadmap around four pivotal tasks — 1. building the update server (software supply chain), 2. customizing the Android system, 3. implementing expanded Open-TEE-based TAs (Trusted Applications), and 4. conducting integrated tests after all builds.

\subsection{Update Server (Supply Chain) Implementation}

\subsubsection{Goal}
Establish an \textbf{end-to-end trust chain} that can safely package and distribute blockchain modules and Android partition images.  
Secure \textbf{transparency and auditability} that can fundamentally block malicious module injection, version spoofing, and rollback attacks.

\subsubsection{Main Tasks}
\begin{enumerate}
    \item \textbf{CI/CD pipeline construction}\\
    \textit{Automate} static analysis, dynamic instrumentation, and internal penetration testing in pipelines such as GitLab CI or GitHub Actions.  
    Formalize \textit{chain-specific lint rules} and \textit{side-channel patterns} as rule sets, scanning every pull request stage.
    
    \item \textbf{OTA signing and version management}\\
    \textbf{Two-stage signing scheme}: (i) the module developer’s PGP signature, (ii) the central registry’s RSA-PSS signature.  
    Use a transparency log (e.g., Sigstore rekor) to notarize “who, when, what” was distributed.  
    Defend downgrade attacks with \textit{semantic versioning + rollback index}.
    
    \item \textbf{Integrity verification and content delivery}\\
    Adopt a CDN + edge cache structure, while the client re-verifies \textbf{SHA-256/SHA-3 hashes}.  
    Compress transfer volume (\textless{} 30\%) and shorten application time (\textless{} 10~s \texttt{@} UFS~3.1) with a delta-OTA generator.
    
    \item \textbf{Role-based access control}\\
    Separate three roles — administrator, validator, contributor (module developer) — with RBAC.  
    Final distribution requires “approval by at least two validators + automatic test pass”.
\end{enumerate}

\subsubsection{Expected Challenges and Responses}
\begin{itemize}
    \item \textit{Vulnerability hype cycle}: Rapidly rotate rule sets and test benches to respond to new supply-chain attack techniques (e.g., DLL planting, cross-compile backdoors).
    \item \textit{Offline signing device management}: Complete a key rotation policy and monitor with transparency logs in the event of developer HSM token theft or loss.
\end{itemize}

\subsection{Android Customizing}

\subsubsection{Goal}
Mount a \textbf{wallet-specialized runtime service} that supports TrustZone SMC calls and module installation.  
Minimize damage to system security (SELinux, Verified Boot) and user experience (UI/UX).

\subsubsection{Main Tasks}
\begin{enumerate}
    \item \textbf{Wallet system service (“Wallet Manager”)}\\
    Add a HAL (or Binder interface) between the Java framework layer and the native system server.  
    Store module states (installation, version, integrity) in system properties + an SQLite DB and manage their life cycle.
    
    \item \textbf{SELinux policy extension}\\
    Create \texttt{tee\_exec}, \texttt{fota\_update}, and \texttt{wallet\_service} domains and granularly allocate privileges under the \textit{least-privilege} principle.  
    Tune macros and permissions to pass CTS / STS / VTS.
    
    \item \textbf{Boot / DM-Verity integration}\\
    Dynamically mount a \textbf{module partition (virtual APEX)} in an \textit{A/B partition} structure and register its hash tree in the boot verifier.
    
    \item \textbf{UX and internationalization}\\
    Separate multilingual (i18n) string resources and display module download progress and verification status in real time.  
    Secure compatibility with Android Accessibility Service (TalkBack, screen reader).
\end{enumerate}

\subsubsection{Expected Challenges and Responses}
\begin{itemize}
    \item \textbf{Compatibility variance with existing OEM firmware}: Gradually port from a Treble GSI-based reference device (e.g., Pixel~6) to multiple chipsets (Exynos, Snapdragon, MediaTek).
    \item Potential breakage of module ABI compatibility with kernel update cycles $\rightarrow$ write \textit{compat shims} and automate testing.
\end{itemize}

\subsection{OPEN-TEE TA Implementation}

\subsubsection{Goal}
Provide a TA skeleton that can instantly add major and minor chains in \textbf{plug-in form}.  
Minimize build complexity with \textbf{E-development / R-runtime separation} (dependency injection) on the same code base.

\subsubsection{Main Tasks}
\begin{enumerate}
    \item \textbf{TA SDK and template}\\
    Modularize standard layers such as \texttt{cryptolib.h}, \texttt{bip32.h}, and \texttt{txserializer.h}.  
    Automatically generate the basic structure (\texttt{tee\_entry.c}, \texttt{param.c}, \texttt{crypto.c}) with a code generator (CLI) by inputting a “foo-chain” schema.
    
    \item \textbf{Chain-specific crypto engine}\\
    Design a virtual function table (VFT) instead of \texttt{\#ifdef} for secp256k1 / ed25519 / sr25519.  
    Performance target for key derivation, signing, and validation: within 1~ms (256-bit ECDSA) \texttt{@} Cortex-A78~2.8 GHz.
    
    \item \textbf{Memory and timing side-channel mitigation}\\
    Implement \textbf{DPA-resistant scalar multiplication} (\texttt{fixed-window + constant-time}).  
    Enable \texttt{TEE\_Malloc} zeroization, stack canary, L1 cache preload, and partition.
    
    \item \textbf{Cross-module IPC}\\
    Use the GP TEE internal core’s \textit{message forwarder} instead of direct TA-to-TA calls, and control session keys with an Argon2 KDF.
\end{enumerate}

\subsubsection{4.4.3 Expected Challenges and Responses}
\begin{itemize}
    \item Non-standard signature schemes of niche chains (e.g., BLS12-381 fast aggregate) $\rightarrow$ need to optimize multiple-precision multiplication in the built-in library.
    \item Efficient cache strategy required when deriving deep HD wallets (BIP-44) under a memory constraint (1 MB TA heap limit).
\end{itemize}

\subsection{Entire Build and Test}

\subsubsection{Goal}
\textbf{Fully automate} functional, security, performance, and compatibility tests in the CI/CD stage.  
Secure deployment trust with \textbf{field-scenario-based} verification, not just laboratory-level testing.

\subsubsection{Test-suite Design}
\begin{enumerate}
    \item \textbf{Function (unit \& integration)}\\
    Achieve 100\% coverage for normal and abnormal input scenarios on each CA·TA IPC path using Google Test / Catch2.  
    Stress-test multi-module simultaneous calls (simultaneous 50 TPS).
    
    \item \textbf{Security (penetration test \& fuzzing)}\\
    Perform dumb fuzzing with AFL++ / libFuzzer and grammar-aware fuzzing (raw transaction hex).  
    Conduct side-channel analysis by collecting event counters (PMU) with \texttt{arm-hpc-toolkit} and applying t-test and TVLA.
    
    \item \textbf{Performance (benchmark)}\\
    Measure signature latency, module load time, OTA application time, and battery consumption (mWh).  
    Target: average 1.5 ms $\pm$ 0.2 ms for signing five chains concurrently, OTA (10 MB) \textless{} 15 s \texttt{@} Wi-Fi 6.
    
    \item \textbf{Compatibility (device farm)}\\
    Use Firebase Test Lab and AWS Device Farm to test 40 devices including custom ROMs such as Android 11–15/One UI, MIUI.  
    Measure physical devices (bare metal) for real network and power profiles, not just emulators.
\end{enumerate}

\subsubsection{Result Analysis and Continuous Improvement}
Aggregate test results in an \texttt{allure-report}; if critical regressions or vulnerabilities are found, issue a GitLab ticket and send a Slack webhook alert.  
“Bug severity → hotfix SLA” matrix: Critical 24 h, High 72 h, Medium 7 d, Low 30 d.  
Comply with the \textbf{coordinated vulnerability disclosure (CVD)} process for CVEs; vendor patch diffs and PoCs are published 90 days later.

\subsection{Long-term Research Extension Directions}
\begin{enumerate}
    \item \textbf{Cross-device key-migration protocol}\\
    Exchange session keys with QR-code-based OOB channels + P256 ECDH, and implement multi-device recovery using Shamir secret sharing.
    
    \item \textbf{Confidential computing integration}\\
    Compare performance and security with ARM CCA (Confidential Compute Architecture) and Intel TDX, and draft a multi-TEE orchestration specification.
    
    \item \textbf{On-device formal verification}\\
    Prototype a \texttt{TEE Proof Assistant} that links to the Tamarin prover and Coq, and automatically prove safety invariants of transaction signatures.
\end{enumerate}

\subsection{Conclusion}
By systematically accomplishing the four practical tasks outlined above, the modular wallet platform proposed by this study can establish itself not merely as a simple prototype but as a reference architecture that simultaneously satisfies safety, flexibility, and operational convenience even in large-scale environments. In particular, securing supply-chain transparency and automating the TA skeleton will be the key to allowing minor chains to enjoy the same security level as the mainstream ecosystem, which is expected to catalyze the evolution of the digital-asset management paradigm into “wallet = platform service.”

\bibliographystyle{IEEEtran}   
\bibliography{references}      

\begin{thebibliography}{10}
\providecommand{\url}[1]{#1}
\csname url@samestyle\endcsname
\providecommand{\newblock}{\relax}
\providecommand{\bibinfo}[2]{#2}
\providecommand{\BIBentrySTDinterwordspacing}{\spaceskip=0pt\relax}
\providecommand{\BIBentryALTinterwordstretchfactor}{4}
\providecommand{\BIBentryALTinterwordspacing}{\spaceskip=\fontdimen2\font plus
\BIBentryALTinterwordstretchfactor\fontdimen3\font minus \fontdimen4\font\relax}
\providecommand{\BIBforeignlanguage}[2]{{%
\expandafter\ifx\csname l@#1\endcsname\relax
\typeout{** WARNING: IEEEtran.bst: No hyphenation pattern has been}%
\typeout{** loaded for the language `#1'. Using the pattern for}%
\typeout{** the default language instead.}%
\else
\language=\csname l@#1\endcsname
\fi
#2}}
\providecommand{\BIBdecl}{\relax}
\BIBdecl

\bibitem{intro_1}
\BIBentryALTinterwordspacing
{Crystal Intelligence}, ``The 10 biggest crypto hacks in history,'' 2025, accessed: 2025-06-22. [Online]. Available: \url{https://crystalintelligence.com/investigations/the-10-biggest-crypto-hacks-in-history/}
\BIBentrySTDinterwordspacing

\bibitem{intro_6}
R.~Mazza, ``Bybit’s 1.4 billion hack: The race to recover stolen crypto from lazarus group,'' \url{https://www.fintechweekly.com/magazine/articles/bybit-ceo-ben-zhou-says-most-funds-are-still-traceable}, Mar. 2025, accessed: 2025-06-22.

\bibitem{intro_2}
Reuters, ``Crypto's biggest hacks and heists after 1.5 billion theft from bybit,'' \url{https://www.reuters.com/technology/cybersecurity/cryptos-biggest-hacks-heists-after-15-billion-theft-bybit-2025-02-24/}, Feb. 2025, accessed: 2025-06-22.

\bibitem{intro_5}
D.~Chavda, ``Dmm bitcoin ends operations after hack, transfers funds to sbi vc,'' \url{https://www.cryptotimes.io/2024/12/02/dmm-bitcoin-to-wind-down-operations-following-305m-hack/}, Dec. 2024, accessed: 2025-06-22.

\bibitem{intro_4}
{The Washington Post}, ``Ftx reports 415 million in hacked crypto, bankman‑fried says it's misleading,'' \url{https://www.washingtonpost.com/technology/2022/11/12/ftx-crypto-hack/}, Nov. 2022, accessed: 2025-06-22.

\bibitem{intro_3}
{BBC News}, ``India fake news: Facebook-whatsapp jail warning as lynchings rise in india,'' \url{https://www.bbc.com/news/world-asia-42845505}, Aug. 2018, accessed: 2025-06-22.

\bibitem{chainalysis2024stolen}
\BIBentryALTinterwordspacing
C.~Team, ``2.2 billion stolen from crypto platforms in 2024,'' Blog post on Chainalysis, Dec. 2024, accessed 2025-06-22. [Online]. Available: \url{https://www.chainalysis.com/blog/crypto-hacking-stolen-funds-2025/}
\BIBentrySTDinterwordspacing

\bibitem{chainalysis2025crime}
\BIBentryALTinterwordspacing
------, ``2025 crypto crime trends from chainalysis,'' Blog post on Chainalysis, Feb. 2025, accessed 2025-06-22. [Online]. Available: \url{https://www.chainalysis.com/blog/2025-crypto-crime-report-introduction/}
\BIBentrySTDinterwordspacing

\bibitem{newstimes2025chetal}
\BIBentryALTinterwordspacing
Newstimes, ``Chetal: Bitcoin \& crypto theft in danbury kidnapping,'' Online news article, 2025, accessed 2025-06-22. [Online]. Available: \url{https://www.newstimes.com/news/article/chetal-bitcoin-crypto-theft-danbury-kidnapping-20387392.php}
\BIBentrySTDinterwordspacing

\bibitem{cointelegraph2024multisig}
\BIBentryALTinterwordspacing
Cointelegraph, ``Multisig \& cold wallets — how secure are they really?'' Online news article, 2024, accessed 2025-06-22. [Online]. Available: \url{https://cointelegraph.com/explained/multisig-cold-wallets-how-secure-are-they-really}
\BIBentrySTDinterwordspacing

\bibitem{cointelegraph2024hacks}
\BIBentryALTinterwordspacing
------, ``Multisig \& cold wallets — how secure are they really?'' Online news article, 2024, accessed 2025-06-22. [Online]. Available: \url{https://cointelegraph.com/explained/multisig-cold-wallets-how-secure-are-they-really}
\BIBentrySTDinterwordspacing

\bibitem{paju2023soktee}
A.~Paju, M.~O. Javed, J.~Nurmi, and J.~Savim, ``Sok: A systematic review of tee usage for developing trusted applications,'' in \emph{ARES 2023 – 18th International Conference on Availability, Reliability and Security, Proceedings}.\hskip 1em plus 0.5em minus 0.4em\relax ACM, Aug. 2023, accessed 2025-06-22.

\bibitem{tripathi2023comprehensive}
G.~Tripathi, M.~A. Ahad, and G.~Casalino, ``A comprehensive review of blockchain technology: Underlying principles and historical background with future challenges,'' \emph{Decision Analytics Journal}, vol.~9, no.~1, p. 100344, 2023.

\bibitem{zheng2017overview}
Z.~Zheng, S.~Xie, H.~Dai, X.~Chen, and H.~Wang, ``An overview of blockchain technology: architecture, consensus, and future trends,'' in \emph{Proc.\ IEEE 6th Int. Congress on Big Data}, 2017, pp. 557--564.

\bibitem{nakamoto2008bitcoin}
S.~Nakamoto, ``Bitcoin: A peer‑to‑peer electronic cash system,'' \emph{White paper}, 2008, online; Bitcoin.org.

\bibitem{costan2016sgx}
V.~Costan and S.~Devadas, ``Intel sgx explained,'' IACR Cryptology ePrint Archive, pp. 1--118, 2016.

\bibitem{mckeen2013hasp}
F.~McKeen, I.~Alexandrovich, A.~Berenzon, C.~V. Rozas, H.~Shafi, V.~Shanbhogue, and U.~R. Savagaonkar, ``Innovative instructions and software model for isolated execution,'' in \emph{Proc. 2nd Workshop on Hardware and Architectural Support for Security and Privacy (HASP)}, 2013, p.~10.

\bibitem{arm2009trustzone}
{ARM Ltd.}, ``Arm security technology: Building a secure system using trustzone technology,'' White Paper, 2009.

\bibitem{kaplan2021memoryencryption}
D.~Kaplan, J.~Powell, and T.~Woller, ``Amd memory encryption,'' AMD White Paper, Oct. 2021.

\bibitem{amd2020sev}
{AMD}, ``Sev-snp: Strengthening vm isolation with integrity protection and more,'' White Paper, 2020.

\bibitem{aosp2025abupdate}
{Android Open Source Project}, ``A/b (seamless) system updates,'' Android Developer Documentation, 2025, \url{https://source.android.com/docs/core/ota/ab}.

\bibitem{paper_1}
\BIBentryALTinterwordspacing
B.~McGillion, T.~Dettenborn, T.~Nyman, and N.~Asokan, ``Open-tee - an open virtual trusted execution environment,'' in \emph{14th IEEE International Conference on Trust, Security and Privacy in Computing and Communications (TrustCom)}, Helsinki, Finland, 2015, author's version of the article to appear in TrustCom 2015, August 20-22, 2015. [Online]. Available: \url{https://open-tee.github.io}
\BIBentrySTDinterwordspacing

\bibitem{paper_2}
\BIBentryALTinterwordspacing
W.~Dai, J.~Deng, Q.~Wang, C.~Cui, D.~Zou, and H.~Jin, ``Sblwt: A secure blockchain lightweight wallet based on trustzone,'' \emph{IEEE Access}, vol.~6, pp. 40\,638--40\,648, 2018. [Online]. Available: \url{https://doi.org/10.1109/ACCESS.2018.2856864}
\BIBentrySTDinterwordspacing

\bibitem{paper_3}
\BIBentryALTinterwordspacing
S.~Houy, P.~Schmid, and A.~Bartel, ``Security aspects of cryptocurrency wallets---a systematic literature review,'' \emph{ACM Computing Surveys}, vol.~56, no.~1, pp. 4:1--4:31, 2023. [Online]. Available: \url{https://doi.org/10.1145/3596906}
\BIBentrySTDinterwordspacing

\bibitem{paper_4}
\BIBentryALTinterwordspacing
``Op-tee documentation,'' Linaro \& OP-TEE Project, Tech. Rep., 2023, version retrieved from https://optee.readthedocs.io/en/latest/. [Online]. Available: \url{https://optee.readthedocs.io/en/latest/}
\BIBentrySTDinterwordspacing

\bibitem{paper_5}
\BIBentryALTinterwordspacing
M.~Gentilal, P.~Martins, and L.~Sousa, ``Trustzone-backed bitcoin wallet,'' in \emph{Proceedings of CS2 2017 (International Workshop on Cryptocurrencies and Blockchain Technology)}.\hskip 1em plus 0.5em minus 0.4em\relax Stockholm, Sweden: ACM, 2017. [Online]. Available: \url{http://dx.doi.org/10.1145/3031836.3031841}
\BIBentrySTDinterwordspacing

\bibitem{halderman2008coldboot}
J.~A. Halderman, S.~D. Schoen, N.~Heninger, W.~Clarkson, W.~Paul, J.~A. Calandrino, A.~J. Feldman, J.~Appelbaum, and E.~W. Felten, ``Lest we remember: Cold boot attacks on encryption keys,'' in \emph{Proc. 17th USENIX Security Symposium}, 2008.

\bibitem{hu2021wallitiq}
Y.~Hu, J.~Park, J.~Byun, and H.~Lee, ``Wallitiq: Smart wallet threat analysis and design,'' in \emph{Proc. ACM CODASPY}, 2021.

\bibitem{byun2024electronics}
J.~Byun, J.~Park, and H.~Lee, ``A secure mobile cryptocurrency wallet based on smart threat analysis,'' \emph{Electronics}, vol.~13, no.~13, 2024.

\bibitem{wallitiq2023whitepaper}
WallitIQ, ``Wallitiq whitepaper,'' 2023.

\bibitem{leguesse2020wallets}
B.~Leguesse, K.~Li, and W.~Enck, ``An analysis of android cryptocurrency wallet apps,'' in \emph{Proc. DIMVA Workshop}, 2020.

\bibitem{erinle2025arxiv}
T.~Erinle, J.~Park, and H.~Lee, ``Blockchain wallet threat model and architecture,'' arXiv preprint arXiv:2504.10123, 2025.

\bibitem{techxplore2021wallets}
{TechXplore (MSU)}, ``Msu researchers identify vulnerabilities in crypto wallets,'' 2021, techXplore News.

\bibitem{biryukov2014deanonymization}
A.~Biryukov, D.~Khovratovich, and I.~Pustogarov, ``Deanonymisation of clients in bitcoin p2p network,'' in \emph{Proc. ACM CCS}, 2014.

\bibitem{heilman2015eclipse}
E.~Heilman, A.~Kendler, A.~Zohar, and S.~Goldberg, ``Eclipse attacks on bitcoin’s peer-to-peer network,'' in \emph{Proc. USENIX Security Symposium}, 2015.

\bibitem{apostolaki2017hijacking}
M.~Apostolaki, A.~Zohar, and L.~Vanbever, ``Hijacking bitcoin: Routing attacks on cryptocurrencies,'' in \emph{Proc. IEEE Symposium on Security and Privacy (S\&P)}, 2017.

\bibitem{karame2012doublespending}
D.~Karame, E.~Androulaki, and S.~Capkun, ``Double-spending fast payments in bitcoin,'' in \emph{Proc. ACM CCS}, 2012.

\bibitem{mnakamoto2008bitcoin}
S.~Nakamoto, ``Bitcoin: A peer-to-peer electronic cash system,'' 2008, bitcoin whitepaper.

\bibitem{buterin2013ethereum}
V.~Buterin, ``Ethereum: A next-generation smart contract and decentralized application platform,'' 2013, bitcoin Magazine.

\bibitem{vasek2015scams}
M.~Vasek and T.~Moore, ``There’s no free lunch, even using bitcoin: Tracking the popularity and profits of virtual currency scams,'' in \emph{Proc. Financial Cryptography}, 2015.

\bibitem{trdcrft2024wrench}
trdcrft, ``The 5 wrench attack explained: What is it and how to protect yourself,'' 2024, accessed: Jun. 22, 2025, https://trdcrft.com/the-5-wrench-attack-explained/.

\bibitem{genkin2015ecdsa}
D.~Genkin, L.~Pachmanov, I.~Pipman, and E.~Tromer, ``Ecdsa key extraction from mobile devices via nonintrusive physical side channels,'' in \emph{Proc. CHES}, 2015.

\bibitem{halborn2021hardwarewallet}
H.~Security, ``Top threats facing cryptocurrency hardware wallets,'' 2021.

\bibitem{openTEE2025docs}
{Open-TEE Project}, ``Open-tee documentation,'' \url{https://open-tee.github.io/documentation}, 2025, accessed: 2025-06-22.

\bibitem{mm0ck3r2025opentee}
mm0ck3r, ``test-for-opentee: Minimal test setup for open-tee,'' \url{https://github.com/mm0ck3r/test-for-opentee}, 2025, accessed: 2025-06-22.

\bibitem{android2025downloadandbuild}
{Android Open Source Project}, ``Trusty tee — download and build,'' \url{https://source.android.com/docs/security/features/trusty/download-and-build}, 2025, accessed: 2025-06-22.

\bibitem{android2025keystore}
------, ``Android keystore system overview,'' \url{https://source.android.com/docs/security/features/keystore}, 2025, accessed: 2025-06-22.

\end{thebibliography}
\end{document}